\documentclass[chicago]{emulateapj}
\usepackage[section]{placeins}
\usepackage{amsmath}
\usepackage[dvips]{color}

\shorttitle{Spiraling accretion flow in G33.92+0.11}
\shortauthors{Liu et al.}
\slugcomment{ApJ published.}
\begin{document}

\title{ALMA resolves the spiraling accretion flow in the luminous OB cluster forming region G33.92+0.11}

\author{Hauyu Baobab Liu\altaffilmark{1}} \author{Roberto Galv\'{a}n-Madrid\altaffilmark{2}} \author{Izaskun Jim\'{e}nez-Serra\altaffilmark{3}} \author{Carlos Rom\'an-Z\'u\~niga\altaffilmark{4}} \author{Qizhou Zhang\altaffilmark{5}} \author{Zhiyun Li\altaffilmark{6}} \author{Huei-Ru Chen\altaffilmark{1,7}}

\affil{$^{1}$Academia Sinica Institute of Astronomy and Astrophysics, P.O. Box 23-141, Taipei, 106 Taiwan \color{blue}hyliu\color{magenta}@\color{blue}asiaa.sinica.edu.tw\color{black}} 
\affil{$^{2}$Centro de Radioastronom\'{i}a y Astrof\'{i}sica, UNAM, A.P. 3-72, Xangari, Morelia, 58089, Mexico}
\affil{$^{3}$European Southern Observatory, Karl-Schwarzschild-Str. 2, 85748, Garching, Germany}
\affil{$^{4}$Instituto de Astronom\'ia, UNAM. Unidad Acad\'emica en Ensenada. Ensenada 22860, Mexico}
\affil{$^{5}$Harvard-Smithsonian Center for Astrophysics, 60 Garden Street, Cambridge, MA 02138}
\affil{$^{6}$Department of Astronomy, P.O. Box 400325, Charlottesville, VA 22904, USA}
\affil{$^{7}$Institute of Astronomy and Department of Physics, National Tsing Hua University, Hsinchu, Taiwan}

\def\Ms{$M_{\rm *}$}
\def\Lsun{$L_{\odot}$}
\def\Hi{H\,{\sc i}~}				
\def\Ht{H$_2$\,}                    
\def\COto{$^{12}$CO(J=2$\rightarrow$1)\ }
\def\cch{C$_2$H }
\def\c2h2{C$_2$H$_2$}
\def\hcccn{HC$_3$N~}

\def\kms{$~\rm km\,s^{-1}$}
\def\micro{\,$\mu$m}
\def\Msun{$M_{\odot}$}

\begin{abstract}
How rapidly collapsing parsec-scale massive molecular clumps feed high-mass stars, and how they fragment to form OB clusters, have been outstanding questions in the field of star-formation.
In this work, we report the resolved structures and kinematics of the approximately face-on, rotating massive molecular clump, G33.92+0.11. 
Our high resolution Atacama Large Millimeter/submillimeter Array (ALMA) images show that the spiral arm-like gas overdensities form in the eccentric gas accretion streams.
First, we resolved that the dominant part of the $\sim$0.6 pc scale massive molecular clump (3.0$^{+2.8}_{-1.4}$$\cdot$10$^{3}$ $M_{\odot}$) G33.92+0.11\,A is tangled with several 0.5-1 pc size molecular arms spiraling around it, which may be connected further to exterior gas accretion streams.
Within G33.92+0.11\,A, we resolved the $\sim$0.1 pc width gas mini-arms connecting with the two central massive (100-300 $M_{\odot}$) molecular cores.
The kinematics of arms and cores elucidate a coherent accretion flow continuing from large to small scales.
We demonstrate that the large molecular arms are indeed the cradles of dense cores, which are likely current or future sites of high-mass star formation.
Since these deeply embedded massive molecular clumps preferentially form the highest mass stars in the clusters, we argue that dense cores fed by or formed within molecular arms play a key role in making the upper end of the stellar and core mass functions. 
\end{abstract}


\section{Introduction} \label{sec:introduction}
Magnetic field, turbulence, and (proto)stellar feedback are well-known physical mechanisms that regulate molecular cloud fragmentation, and play a crucial role in determining star-forming efficiency and stellar/core mass function (Wang et al. 2010; Chen \& Ostriker 2014; Tan et al. 2014; Van Loo et al. 2014).
Recently, much attention is drawn to the hierarchical cloud contraction regulated by filamentary or sheet-like gas structures (Burkert \& Hartmann 2004; Zhang et al. 2009; Myers 2009, 2011; Schneider et al. 2010; Liu et al. 2012a; Toal{\'a} et al. 2012; Busquet et al. 2013; Dobbs et al. 2013).

On the other hand, for (giant) molecular clouds that are undergoing rapid global collapse (e.g. Liu et al. 2012a; Hartmann et al. 2012; V{\'a}zquez-Semadeni et al. 2007; Howard et al. 2014), the amplified effects of the initial specific angular momentum may lead to a distinctive mode of fragmentation in cloud centers (Keto et al. 1991).
The rotationally supported structures can continue accumulating gas, and can have sufficient time to fragment. 
The arm-like gas local overdensities, which is conducive to fragmentation, may also be shock induced at where the eccentric accretion streams collide with each other.
Indeed, the observed marginally centrifugally supported accretion flows (e.g. Galv{\'a}n-Madrid et al. 2009; Liu et al. 2010; Liu et al. 2013; Cesaroni et 
al. 2011) have argued that the gas dynamics can be dictated by angular momentum in the central $<$1 pc regions of molecular clouds.
Several interesting cases have shown fragmentation in rapidly rotation systems. 
For example, the spatially resolved clumpy gas toroid in the center of the low-mass cluster-forming region L1287 and the center of the intermediate-mass star-forming region NGC6334\,V (Juar\'{e}z et al. in prep.), and the ring-like distribution of the ultra compact (UC) H\textsc{ii} regions (Churchwell 2002) in the central $\sim$2 pc of the Galactic mini-starburst region W49N (Welch et al. 1987; De Pree et al. 1997; Galv{\'a}n-Madrid et al. 2013).
From the observational point of view, two outstanding questions remain: a) How the large scale streams converge onto these rotating systems, and b) how the dense cores form subsequently.

\begin{figure*}[ht]
\hspace{1.2cm}
\begin{tabular}{ p{12cm}  }
\includegraphics[width=14cm]{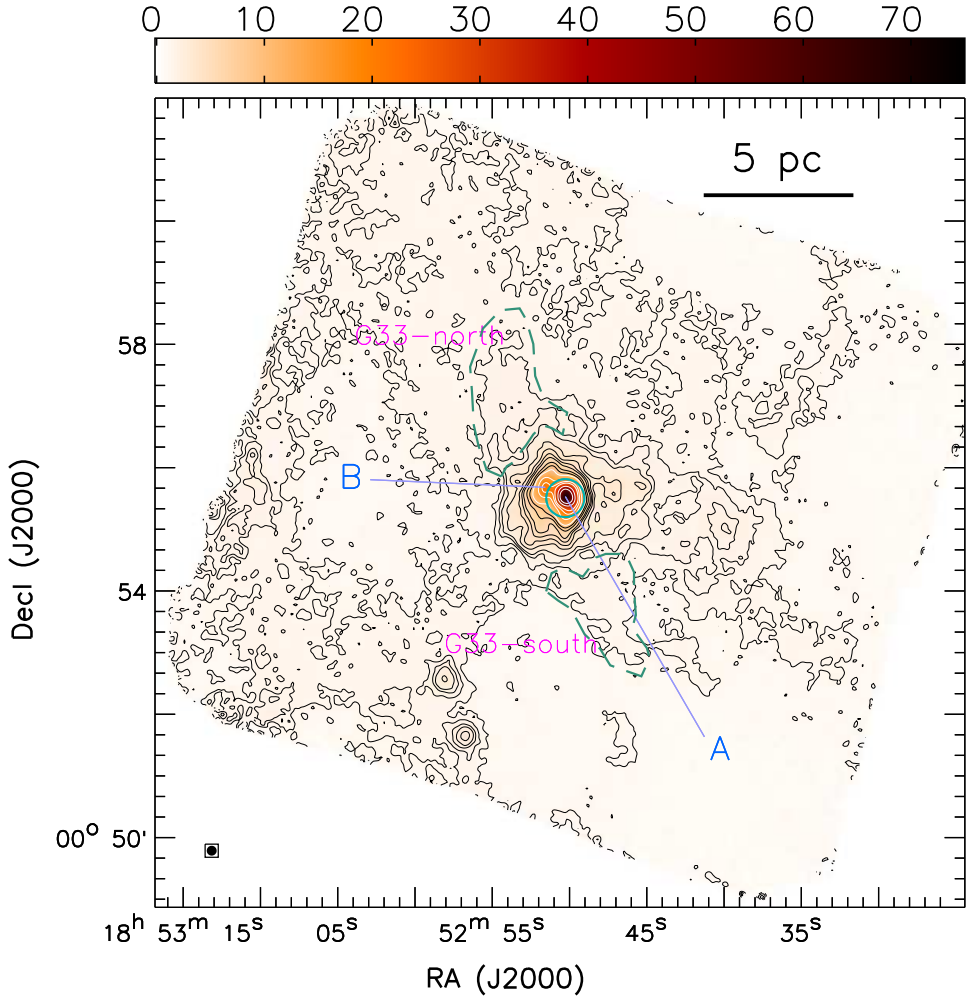}  \\
\end{tabular}
\caption{\footnotesize{The SHARC2/CSO+SPIRE/Herschel 0.35 mm continuum image ($\theta_{\mbox{\scriptsize{HPBW}}}$=9$''$.6). 
The significant contours are not presented with identical intervals, for the sake of tracing the dust/gas column density. Black contours are 250 mJy\,beam$^{-1}$ (5$\sigma$)$\times$[4.5, 6.0, 7.5, 9.0, 12.0, 15.0, 18.0, 21.0, 27.0, 33.0]. 
White contours are 250 mJy\,beam$^{-1}$ (5$\sigma$)$\times$[40, 60, 80, 100, 150, 200]. 
The circle shows the ALMA 12m array primary beamwidth at 15\% power ($\sim$37$''$ in diameter). 
We refer to the two regions enclosed by the green dashed lines as G33-north and G33-south. 
Color bar is in units of Jy\,beam$^{-1}$.
}} 
\vspace{0.8cm}
\label{fig_continuum1}
\end{figure*}

\begin{figure*}[ht]
\hspace{1.2cm}
\begin{tabular}{ p{12cm} }
\includegraphics[width=14cm]{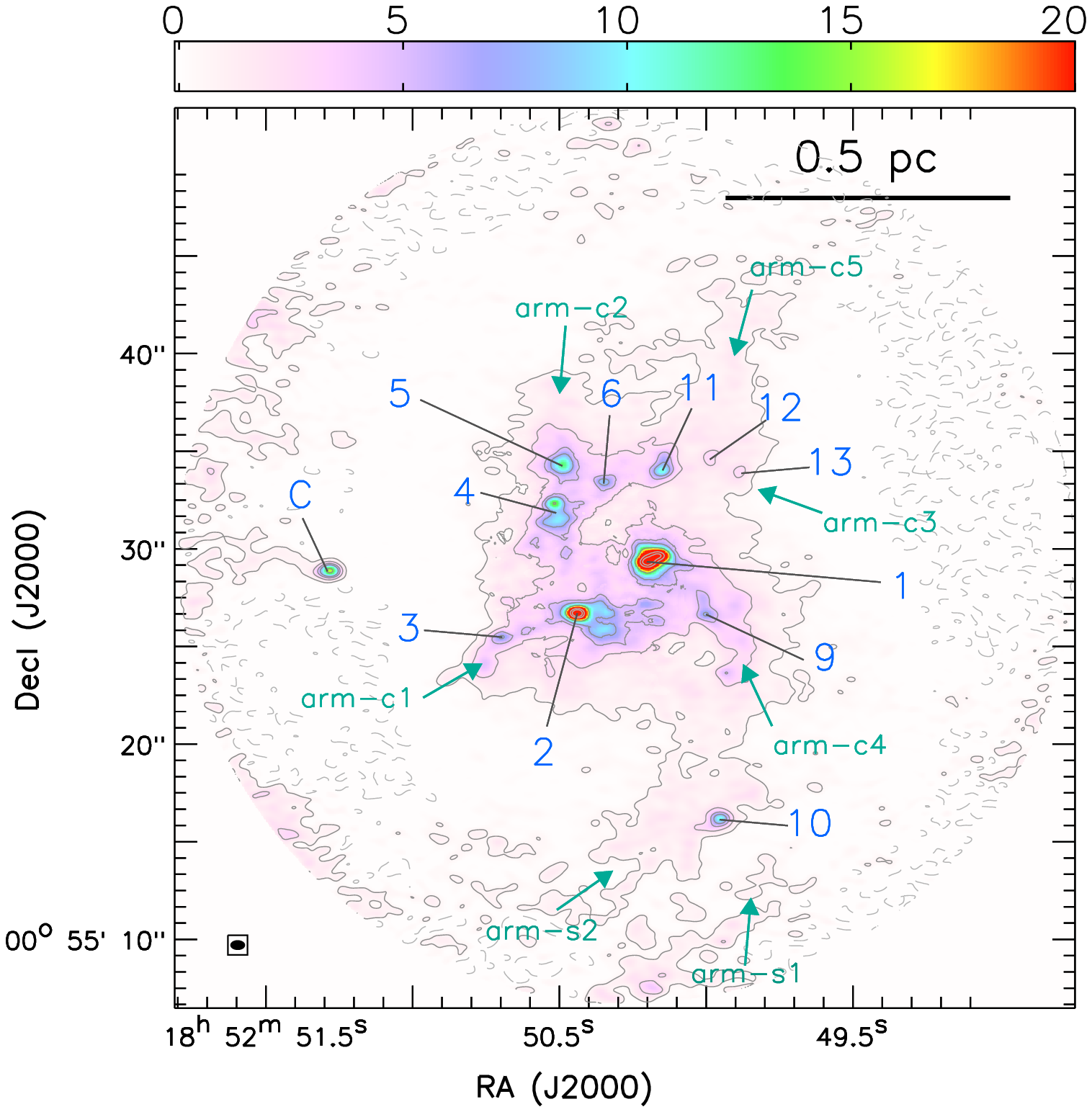} \\
\end{tabular}
\caption{\footnotesize{The free-free model-subtracted dust continuum image, taken by the ALMA 12m+ACA array observations ($\theta_{\mbox{\scriptsize{maj}}}$$\times$$\theta_{\mbox{\scriptsize{min}}}$=0$''$.75$\times$0$''$.50, P.A.=90$^{\circ}$). 
Contour levels are 0.6 mJy\,beam$^{-1}$ (3$\sigma$) $\times$[-1, 1, 4, 8, 12, 24, 48, 60]. 
The peak emission in this image is 46.6 mJy\,beam$^{-1}$ at core A2. 
The color image does not present the full intensity range because the very bright emission regions are compact. 
We label the source C in the east, and label the (A) cores discussed in the text with numbers. 
Arrows indicate the molecular arms.
The color bar is in units of mJy\,beam$^{-1}$.
}} 
\vspace{0.8cm}
\label{fig_continuum2}
\end{figure*}

Since the spinning massive molecular clumps are geometrically flattened, the morphologically well-resolvable cases are limited to those which are far from an edge-on projection.
Their detailed gas kinematics is only now accessible by high spectral resolution observations of molecular line, thanks to the unprecedented sensitivity of the Atacama Large Millimeter Array (ALMA).
We selected to observe the $L_{bol}$$\sim$2.5$\times$10$^{5}$ $L_{\odot}$ OB cluster forming region G33.92+0.11, which is at a distance of 7.1$^{+1.2}_{-1.3}$ kpc (Fish et al. 2003)\footnote{The $v_{lsr}$ of this source is close to the tangent point, and therefore does not have near-far ambiguity in the kinematics distance.}.
Despite the strong free-free emission associated with the already existing embedded OB cluster, our previous Submillimeter Array (SMA) chemical studies suggested that this source is at an early evolutionary stage, and may not yet be seriously disturbed by (proto)stellar feedback (Liu et al. 2012b).
In addition, previous lower angular resolution Berkeley Illinois Maryland Array (BIMA), Very Large Array (VLA), and SMA observations of the molecular line emission  consistently found that the measured molecular gas mass is 10 times larger than the virial mass (Watt \& Mundy 1999; Liu et al. 2012b).
This can be naturally interpreted by the nearly face-on projection. 
Therefore we expect minimized confusion among structures in the line-of-sight.

\begin{figure*}[ht]
\hspace{-0.8cm}
\begin{tabular}{ p{18cm} }
\rotatebox{-90}{
\includegraphics[width=11cm]{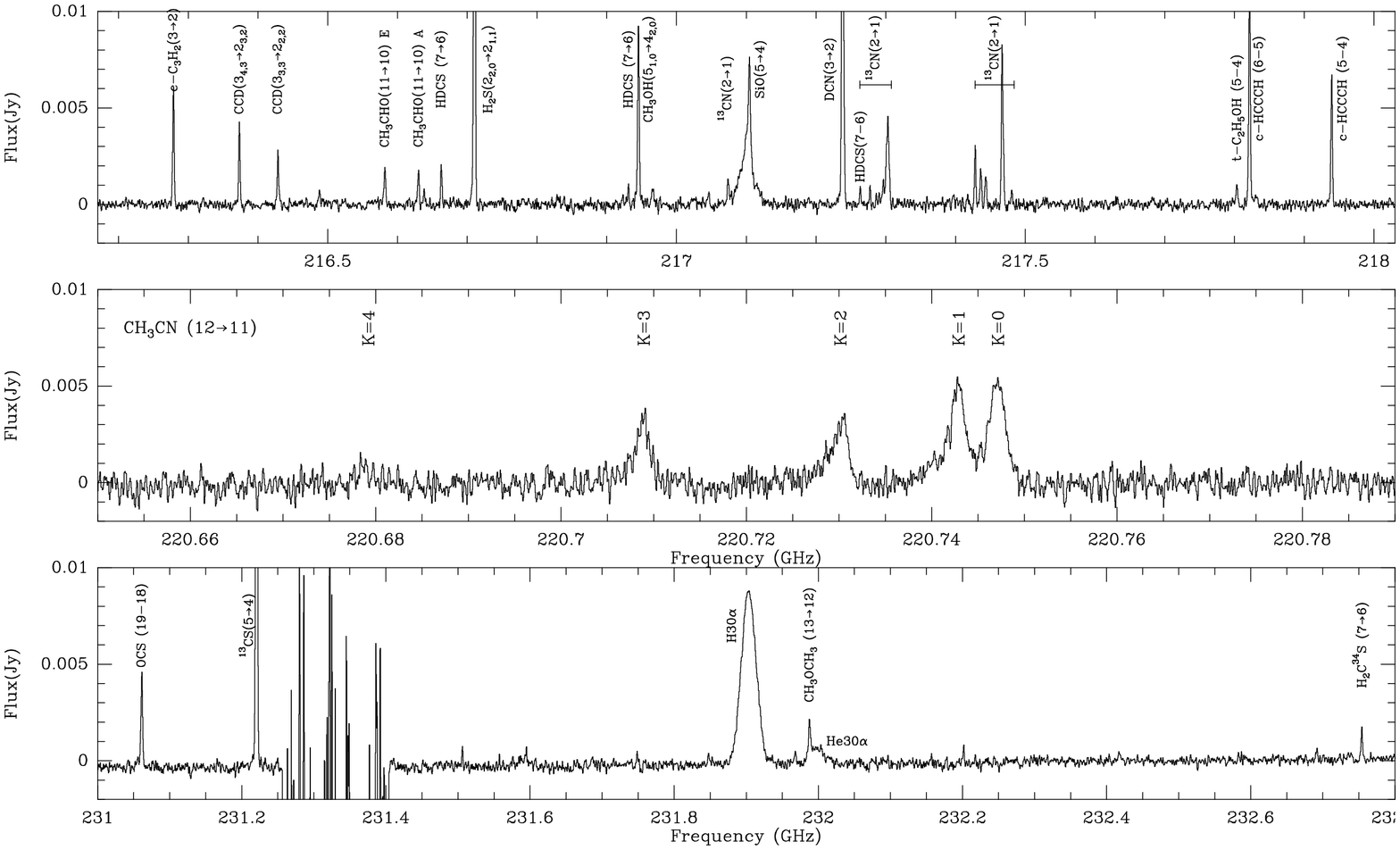}}
\end{tabular}
\caption{\footnotesize{
The spectra taken with the ALMA 12m+ACA Array (Section \ref{subsec:alma}) and the identified spectral lines (see also Table \ref{tab-lines}). Due to bandpass defects, the 12m Array data were lost in 300 spectral channels in the 231 GHz spectral window.
}} 
\vspace{0.5cm}
\label{fig_spectra}
\end{figure*}

The major improvement of the present work over previous observations (e.g. Liu et al. 2012b), is the $\sim$8 times better spectral resolution ($\sim$0.17 km\,s$^{-1}$) in the CH$_{3}$CN J=12-11 observations.
In addition, the unprecedented sensitivity of ALMA allows imaging various optically thiner dense molecular gas tracer at sub-arcsecond resolution. 
This permits discriminating the blended molecular gas arms\footnote{We follow the existing nomenclature in the literature (e.g., Zhang et al. 2009; Liu et al. 2012ab). 
In this way, \textit{Massive molecular clumps} refer to structures with sizes of $\sim$0.5-1 pc, \textit{massive molecular cores} refer to the $<$0.1 pc size structures embedded within a clump, and \textit{condensations} refer to the distinct molecular substructures within a core. 
\textit{Fragmentation} refers to the dynamical process that produces or enhances multiplicity. 
\textit{Molecular filaments} refer to the geometrically elongated molecular structures, and \textit{molecular arms} refer to segments of molecular filaments that are located within the $\lesssim$1 pc radii of molecular clumps and may not be fully embedded within molecular clumps.}, which is important for diagnosing the details of  molecular gas kinematics.
In this paper, we focus on the 1.3 mm dust continuum emission, which is a reliable tracer of the molecular gas column density; the DCN 3-2 line, which was observed  to be less abundant in hot molecular cores (Rodgers \& Millar 1996; Hatchell et al. 1998); the $^{13}$CS 5-4 line which is only excited in warm and dense gas and may be shock-enhanced; and the hot molecular core tracer, the CH$_{3}$CN J=12-11 K-ladders (Sutton et al. 1986;  MacKay 1999; Araya et al. 2005). 
Based on these observations, we are able to further differentiate the kinematics gas streams at different physical conditions, which may be well blended in the spatial or velocity domain. 
Our ALMA observations also cover the recombination lines H30$\alpha$ and He30$\alpha$, which trace the ionized gas.
We additionally present the SHARC2/CSO (Dowell et al. 2003) and SPIRE/Herschel\footnote{Herschel is an ESA space observatory with science instruments provided by European-led Principal Investigator consortia and with important participation from NASA.} (Griffin et al. 2010) observations of the 350 $\mu$m continuum emission, to outline the overall molecular cloud geometry. 
Details of our observations are provided in Section \ref{sec:observation}.
We present the obtained continuum image and the molecular line data cubes in Section \ref{sec:result}.
Section \ref{sub:matter} and \ref{subsec:satellite} will address the observed matter distribution and the gravitational stability of the resolved gas structures. 
Our interpretation of the observed velocity field is provided in Section \ref{sub:velocity}.
The physical implication of our results is discussed in Section \ref{sub:implication}.
A brief summary will be given in Section \ref{sec:summary}.
The identified line species from our ALMA observations are provided in Appendix A.

Based on our IRAM-30m mapping observations of the $^{13}$CO 2-1 and C$^{18}$O 2-1 line (Liu et al. 2012c), we are sure that all gas structures traced by the dust continuum emission addressed in this manuscript are physically associated.
We will present the larger scale gas kinematics traced by the CO 2-1 isotopologues, the details of core mass distribution, the ionized gas kinematics traced by the H30$\alpha$ and He30$\alpha$ lines, as well as the chemistry, in separated papers.

\section{Observations} \label{sec:observation}

\subsection{ALMA 12m+ACA Array 1.3 mm} \label{subsec:alma}
The ALMA 12m Array (i.e. 12m dish size) observations were carried out on 2014 May 04, with $\sim$36 good antennas.
The pointing and phase referencing center is R.A. (J2000) =18$^{\mbox{\scriptsize{h}}}$52$^{\mbox{\scriptsize{m}}}$50$^{\mbox{\scriptsize{s}}}$.272, and decl. (J2000) =00$^{\circ}$55$'$29$''$.604. 
We used two 234.4 MHz wide spectral windows (channel spacing 61 kHz, $\sim$0.085 km\,s$^{-1}$), and two 1875.0 MHz wide spectral windows (channel spacing 488 kHz, $\sim$0.65 km\,s$^{-1}$), which tracked the velocity of $v_{lsr}$$\sim$107.6 km\,s$^{-1}$ on our target source.

The central frequency of these spectral windows are 231.220690 GHz ($^{13}$CS 5-4), 231.900928 GHz (H30$\alpha$), 220.679320 GHz (CH$_{3}$CN J=12-11, K=4), and 217.104980 GHz (SiO 5-4), respectively.
The {\it uv} sampling range of these observations is 13-430 k$\lambda$.
The overall on-source time was 43.7 minutes, and the T$_{sys}$ was in range of 60-150 K.
We observed Titan, J1851+0035, and J1751+0939, for absolute flux, gain, and passband calibrations, respectively.

\begin{figure*}
\vspace{-0.1cm}
\hspace{-0.8cm}
\begin{tabular}{ p{5.9cm} p{5.9cm} p{5.9cm} }
\includegraphics[width=6.4cm]{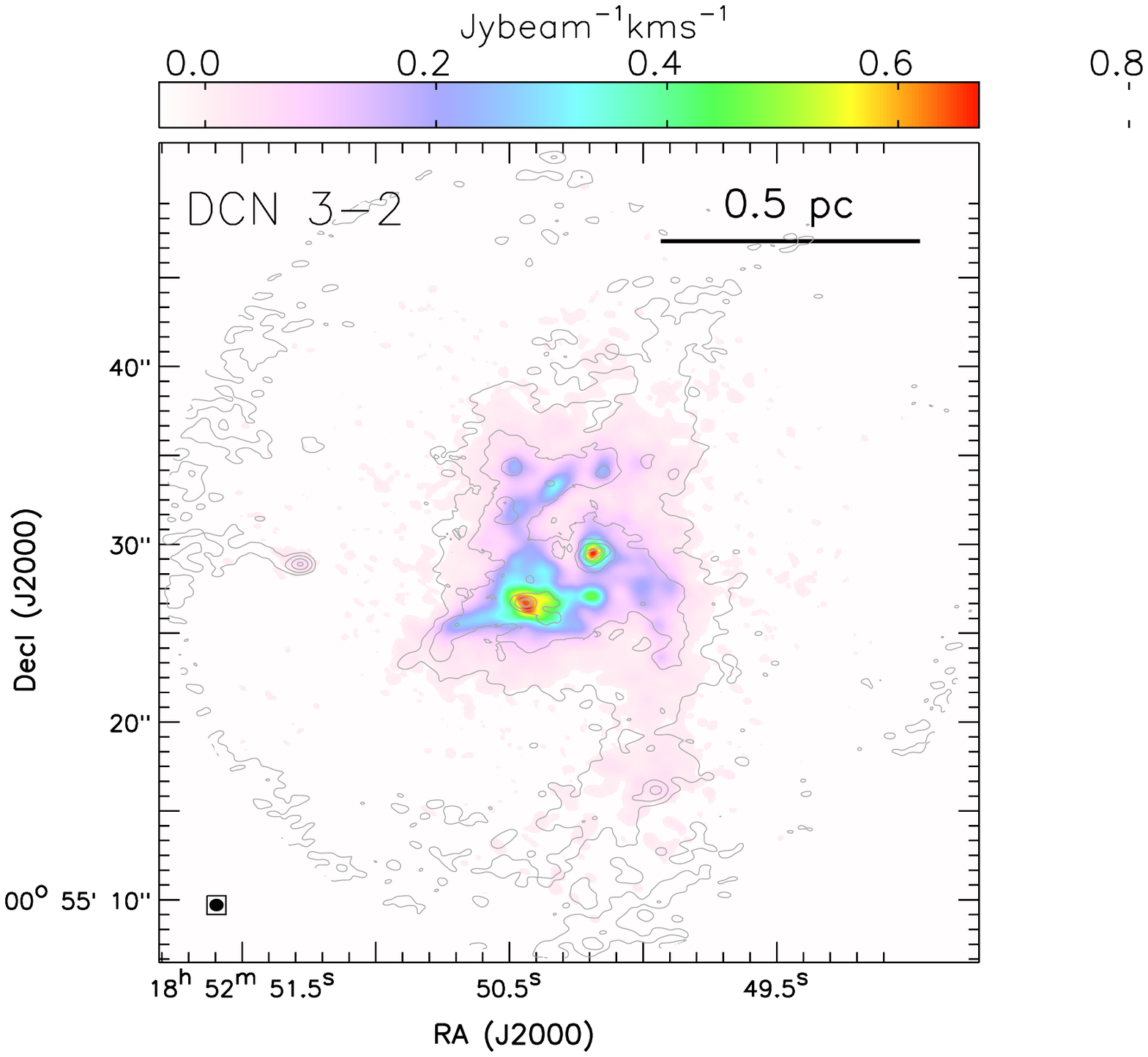} & \includegraphics[width=6.4cm]{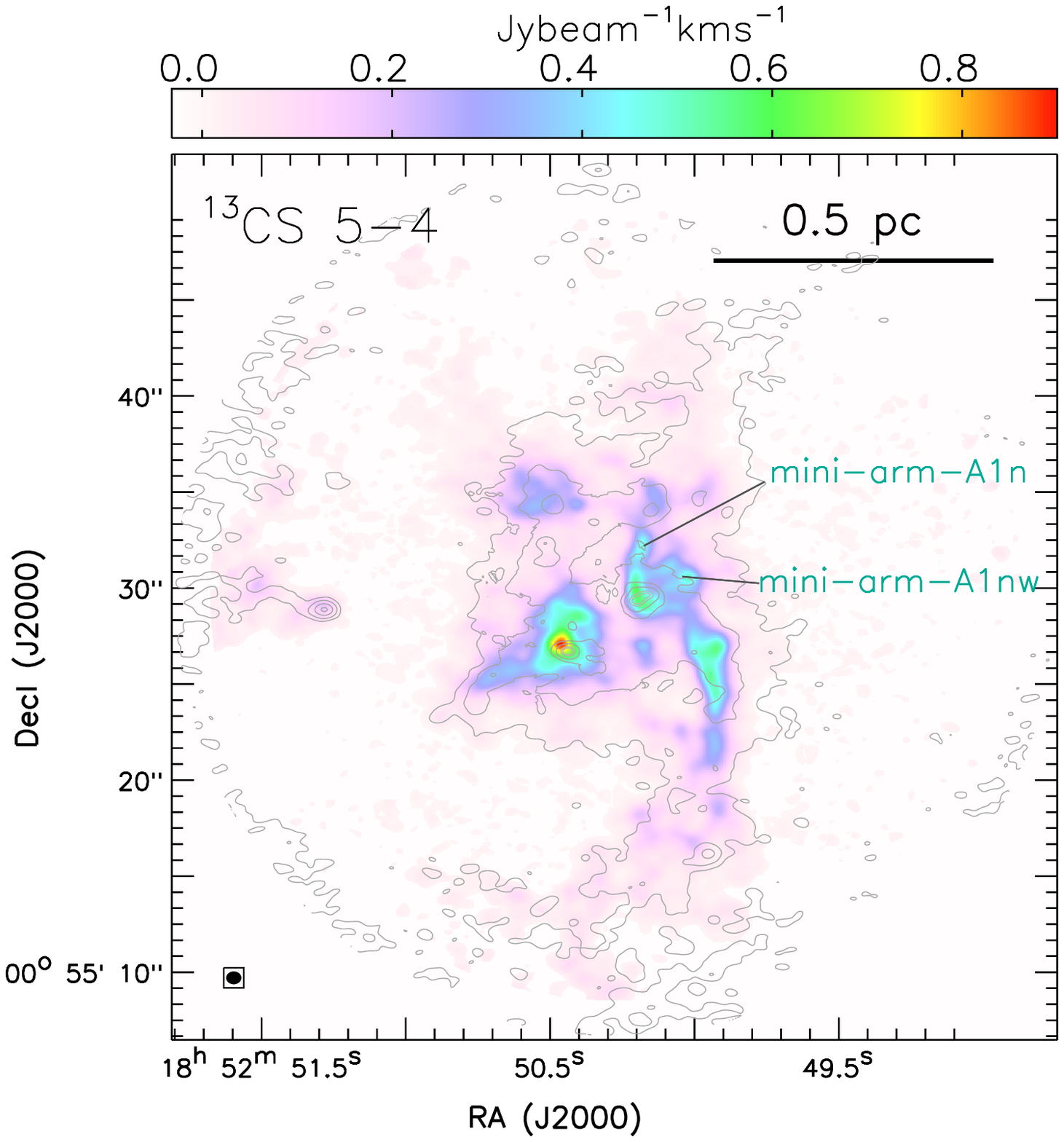} & \includegraphics[width=6.4cm]{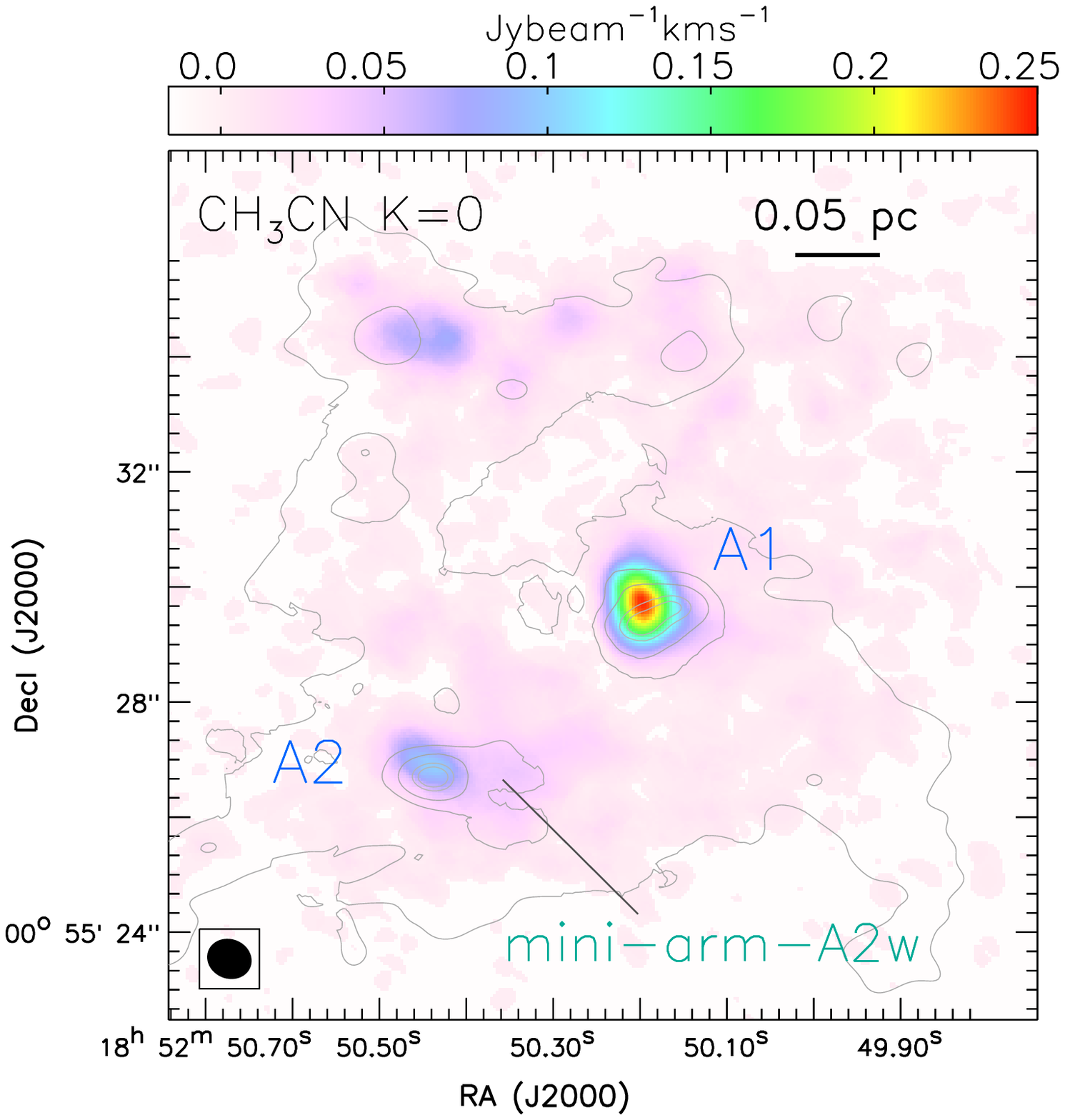}\\ 
\includegraphics[width=6.4cm]{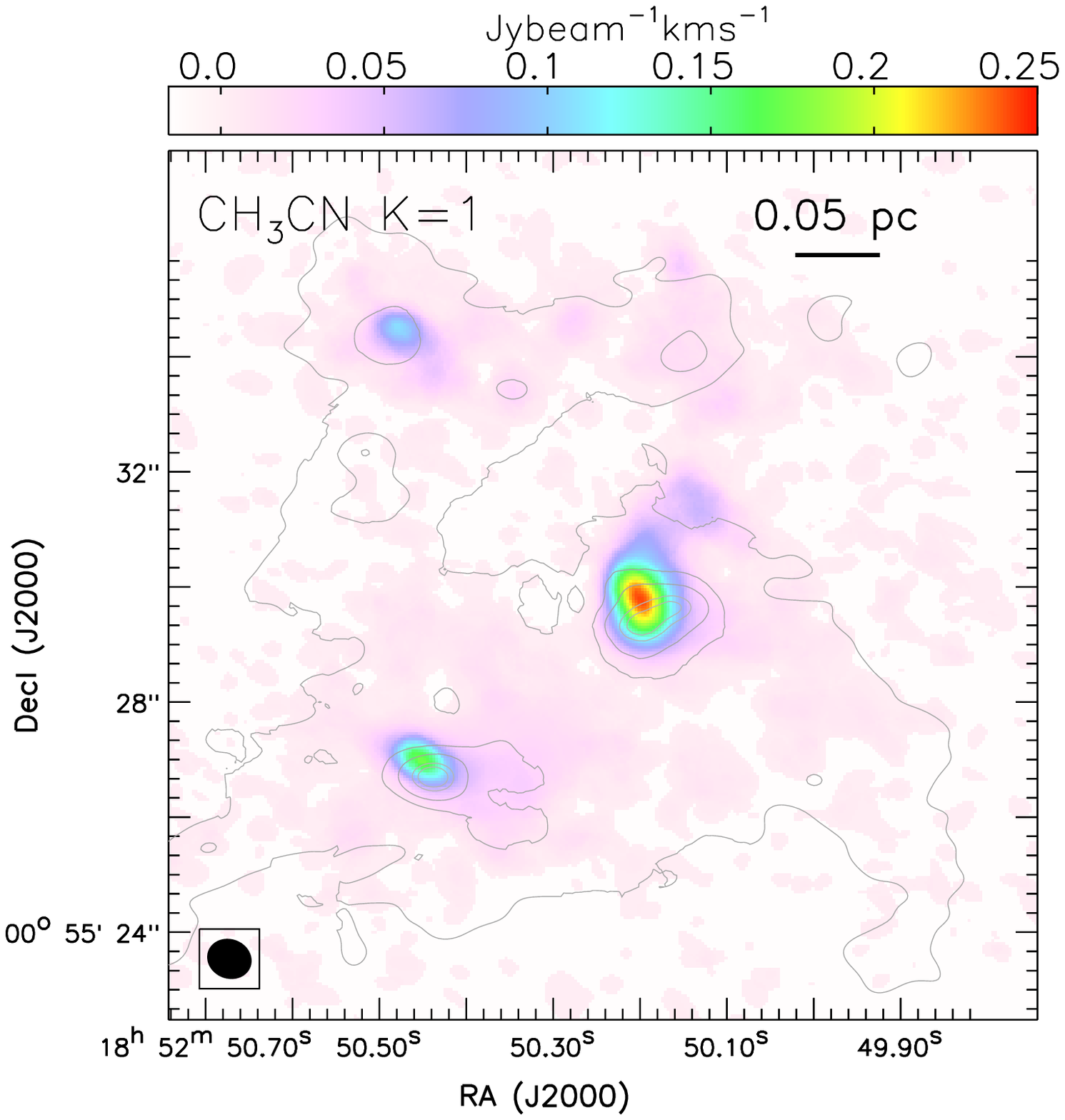} & \includegraphics[width=6.4cm]{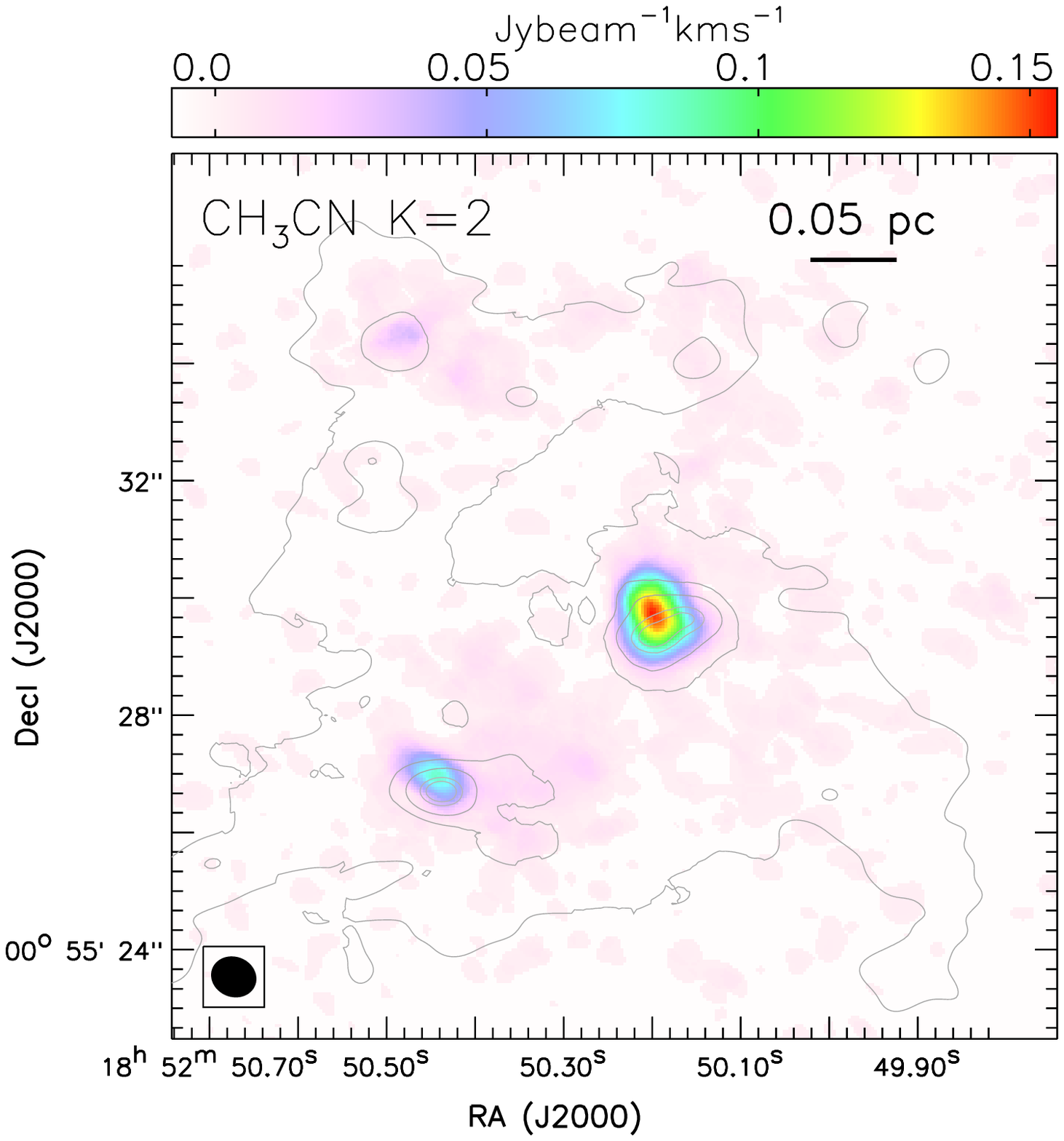} & \includegraphics[width=6.4cm]{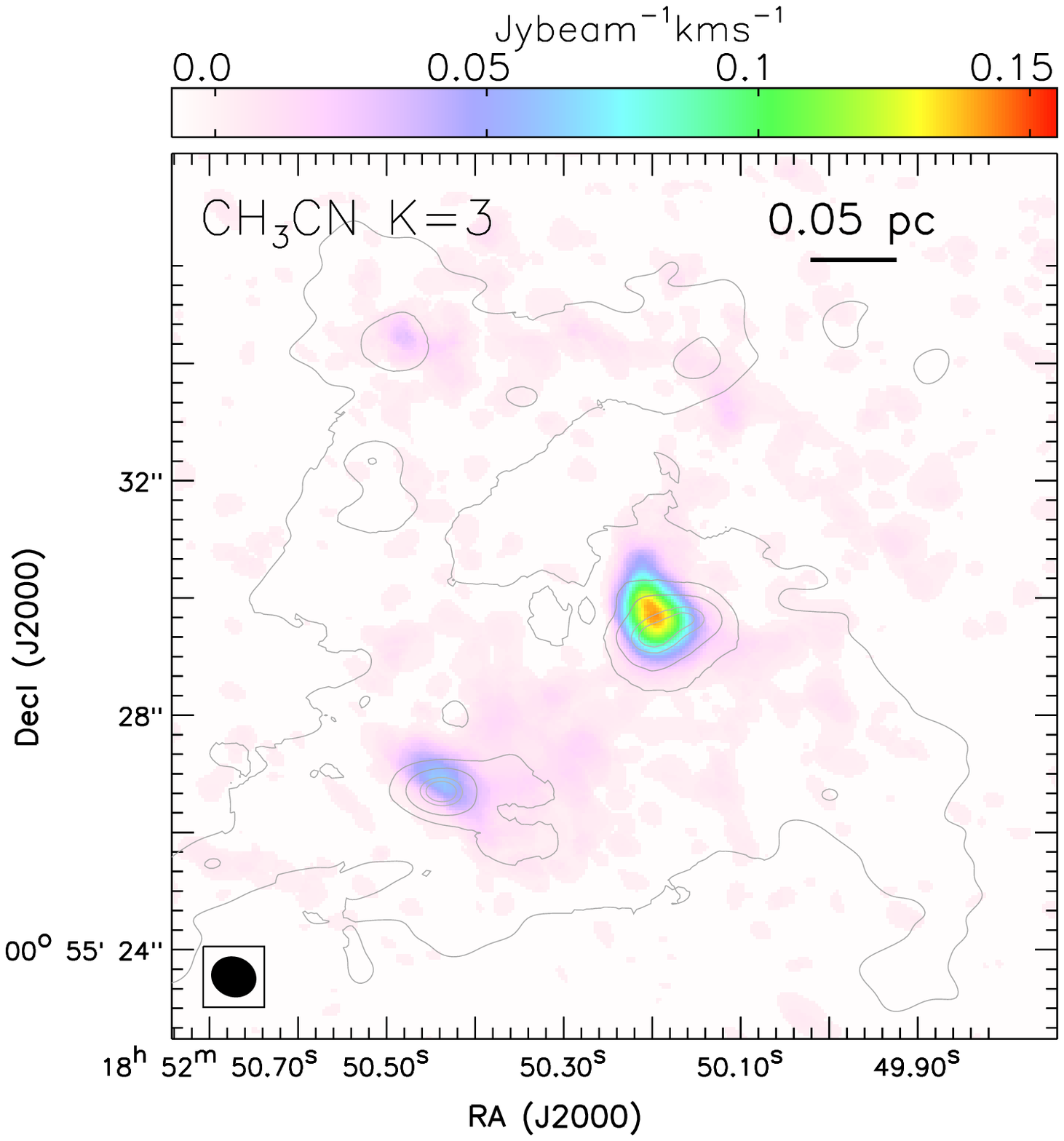}\\ 
\end{tabular}
\caption{\footnotesize{The velocity integrated intensity maps of the selected molecular lines (color), overlaid on the 1.3 mm dust continuum image (contours). Contours are 0.6 mJy\,beam$^{-1}$ (3$\sigma$)$\times$[1, 4, 12, 24, 48, 60] in the $^{13}$CS and DCN line panels, and are 0.6 mJy\,beam$^{-1}$$\times$[4, 12, 24, 48, 60] in the panels for the CH$_{3}$CN K-ladders. 
}} 
\vspace{0.5cm}
\label{fig_moment0}
\end{figure*}

The ALMA Compact Array (ACA; 7m dishes) observations were carried out on 2014 May 03 and 2014 May 04, with $\sim$10 available antennas\footnote{Not all antennas were always valid for the observations because of the antenna shadowing.}.
The pointing and phase referencing center, and the spectral setup, are the same with the 12m array observations. 
These ACA observations cover a {\it uv} sampling range of 7.2-47 k$\lambda$.
The overall on-source time is 24.4 minutes on May 03, and 73.2 minutes on May 04.
The T$_{sys}$ was in range of 60-150 K.
We observed Titan/Neptune, J1851+0035, and J1733-1304/J1924-2914, for absolute flux, gain, and passband calibrations, respectively.

The data were calibrated, phase self-calibrated (2 iterations), and jointly imaged using the CASA package (McMullin et al. 2007) version 4.2.0 (release r28322).
In the end, we did not implement the phase self-calibration solution to the 12m Array data because there was no obvious improvement in image quality.
We did not self-calibrate the visibility amplitude because of a significant missing flux ($\gtrsim$24\%) as compared with previous SMA observations in the same region.
The continuum image was constructed by jointly imaging all spectral line-free channels using the multi-frequency imaging (\texttt{mfs}) algorithm.
The Briggs Robust=0 weighted 12m+ACA continuum image has a synthesized beam of $\theta_{\mbox{\scriptsize{maj}}}$$\times$$\theta_{\mbox{\scriptsize{min}}}$=0$''$.67$\times$0$''$.47 (P.A.=81$^{\circ}$), and the dynamic range limited rms noise level of 0.2 mJy\,beam$^{-1}$.
We have corrected the primary beam attenuation, and therefore the images are noisier in outer region. 

The selected $^{13}$CS 5-4 line was simultaneously covered in one 234.4 MHz spectral window, and a 1875.0 MHz spectral window.
Due to a correlator failure, we lost the $^{13}$CS 5-4 data in the 234.4 MHz spectral window. 
The CH$_{3}$CN J=12-11 K-ladders were observed in a 234.4 MHz spectral window; the DCN 3-2 and SiO 5-4 lines were observed in 1875.0 MHz spectral windows.
We performed Briggs Robust=2 weighted 12m+ACA spectral line imaging to enhance the sensitivity to diffuse emission, which yielded the synthesized beam of $\theta_{\mbox{\scriptsize{maj}}}$$\times$$\theta_{\mbox{\scriptsize{min}}}$=0$''$.78$\times$0$''$.67.
The standard 1-iteration spectral hanning smoothing procedure, which can better suppress the passband ripple, will yield $\sim$0.17 km\,s$^{-1}$ velocity resolution, and $\sim$1.3 km\,s$^{-1}$ velocity resolutions for the 234.4 MHz and the 1875.0 MHz spectral windows, respectively.
However, we imaged the $^{13}$CS 5-4 and the DCN 3-2 lines with $\sim$0.68 km\,s$^{-1}$ channel spacing, intending to better present the kinematics information.
The CH$_{3}$CN J=12-11 K-ladders were imaged with the 0.17 km\,s$^{-1}$ velocity resolution. 
We achieved the rms noise level of 1.6 mJy\,beam$^{-1}$ (70 mK) and 3.5 mJy\,beam$^{-1}$ (170 mK) in the $\sim$0.65 km\,s$^{-1}$ and 0.17 km\,s$^{-1}$ velocity channels, respectively.
There is no zero-spacing information available for our 1.3 mm continuum map.
We have compared our ALMA observations with the existing SMA observations of the 1.3 mm continuum emission, the CSO SHARC2 bolometric single-dish observations of the 0.35 mm continuum emission (see below), and the VLA observations of the NH$_{3}$ lines (Liu et al. 2012b). 
We confirmed that the resolved morphology by ALMA is not significantly biased by the missing flux on the $\lesssim$30$''$ scale. 
The ALMA 12m+ACA Array observations of the dense molecular gas tracers are less biased by missing zero spacing because the emission regions are less extended in the individual velocity channels.
Presentation of the ALMA images will be limited to the 15\% power of the primary beam width, if not specifically mentioned.

\subsection{Herschel SPIRE 0.35 mm} \label{subsec:herschel}
For this study we made use of a 350 $\mu$m map of G33.92+0.11 obtained by the Herschel space telescope. 
We used the level 3 (science grade) processed mosaics of the SPIRE-PACS paralell observations of a much larger map (observation ID 1342207031) of the W49 complex observed in october 2010. 
No significant glitches are visible in the region of interest. 
Level 3 maps are produced from a combination of all available contiguous observations of a program. 
The Level 3 mosaics have optimal WCS solutions and analysis quality grade with zero points calibrations estimated using Planck Obsevatory maps, and archived directly in units of surface brightness (MJy/sr). 
We applied, however, a simple astrometric check by re-creating the mosaic with \texttt{Montage}. \footnote{http://montage.ipac.caltech.edu}, and comparing positions of compact features against a WISE Observatory 3.6 micron map.

\subsection{CSO SHARC2 0.35 mm} \label{subsec:cso}
High angular resolution continuum observations at 0.35 mm were carried out using the SHARC2 bolometer array, installed on the Caltech Submillimeter Observatory (CSO).
The array consists of 12$\times$32 pixels (approximately 85\% of these pixels work well).
The simultaneous field of view (FOV) provided by this array is 2$'$.59$\times$0$'$.97, and the diffraction limited beam size is $\sim$8$''$.8.

The data were acquired on 2014 March 24 ($\tau_{\mbox{\scriptsize{225 GHz}}}$$\sim$0.06).
The telescope pointing and focusing were checked every 1.5-2.5 hours.
Mars was observed for absolute flux calibration.
We used the standard 10$'$$\times$10$'$ on-the-fly (OTF) box scanning pattern, centered on R.A. (J2000) =18$^{\mbox{\scriptsize{h}}}$52$^{\mbox{\scriptsize{m}}}$50$^{\mbox{\scriptsize{s}}}$.272, and decl. (J2000) =00$^{\circ}$55$'$29$''$.604.
The total on-source time was 30 minutes.
Data calibration was performed using the CRUSH software package (Kov\'{a}cs 2008).
In addition, we used the MIRIAD (Sault et al. 1995) task \texttt{immerge} to perform the weighted sum of the SHARC2 0.35 mm image with the Herschel SPIRE image (see Section \ref{subsec:herschel}) at the similar wavelength, in the fourier domain.
This minimizes the missing flux in the ground based single-dish bolometric observations due to the sky subtraction, which typically causes defects on angular scales larger than the simultaneous field of view (or the angular throw in the nodding observations). 
The final map was smoothed to an angular resolution of 9$''$.6, and the rms noise level achieved was $\sim$60 mJy\,beam$^{-1}$.
In addition, we aligned the final 0.35 mm image with the Herschel, ALMA, and SMA images.


\section{Results} \label{sec:result}

\begin{figure}
\vspace{-1.1cm}
\hspace{-2cm}
\begin{tabular}{ p{8.5cm} }
\includegraphics[width=10.5cm]{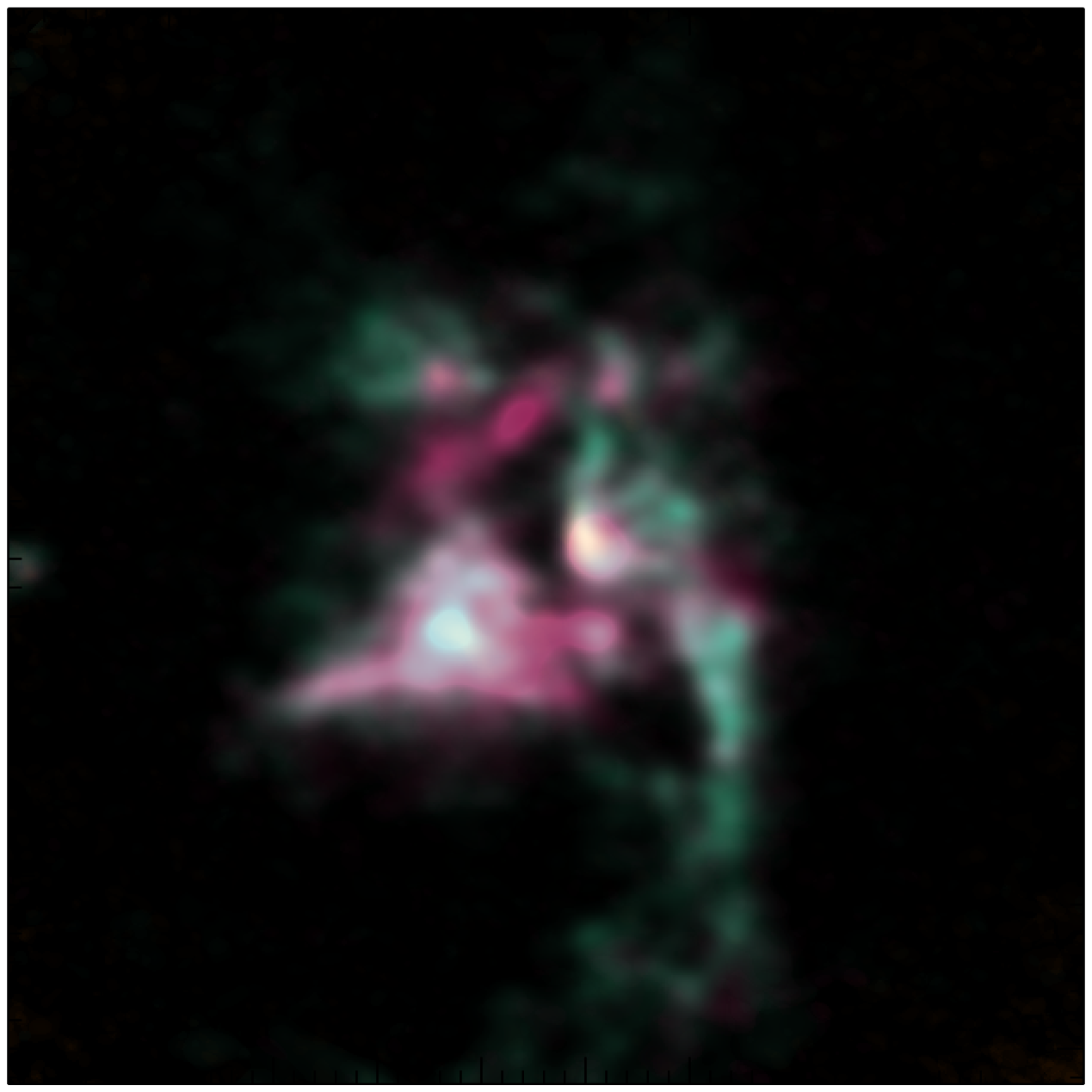} \\
\end{tabular}

\vspace{-2.3cm}
\hspace{-2cm}
\begin{tabular}{ p{8.5cm} }
\includegraphics[width=10.5cm]{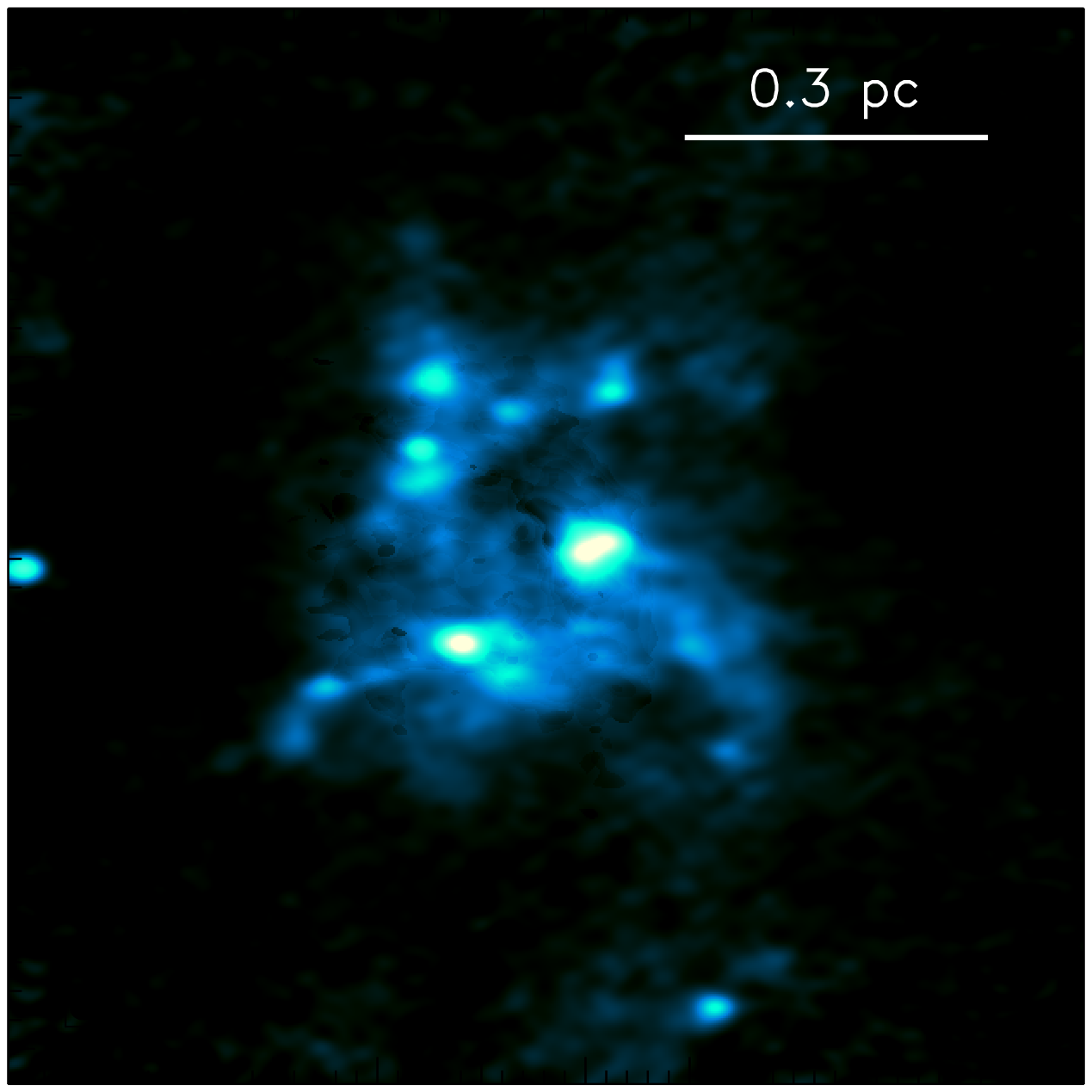} \\
\end{tabular}
\vspace{-1.0cm}
\caption{\footnotesize{Top: The color composited image generated from the integrated intensity maps of DCN 3-2 (magenta), $^{13}$CS 5-4 (cyan), and CH$_{3}$CN J=12-11 K=0 line (yellow). Bottom: The dust continuum image taken by the ALMA 12m+ACA array observations. These two panels are presented on the same spatial scale. 
}} 
\vspace{0.8cm}
\label{fig_rgb}
\end{figure}

\begin{figure*}
\hspace{-0.8cm}
\begin{tabular}{ p{5.9cm} p{5.9cm} p{5.9cm} }
\includegraphics[width=6.4cm]{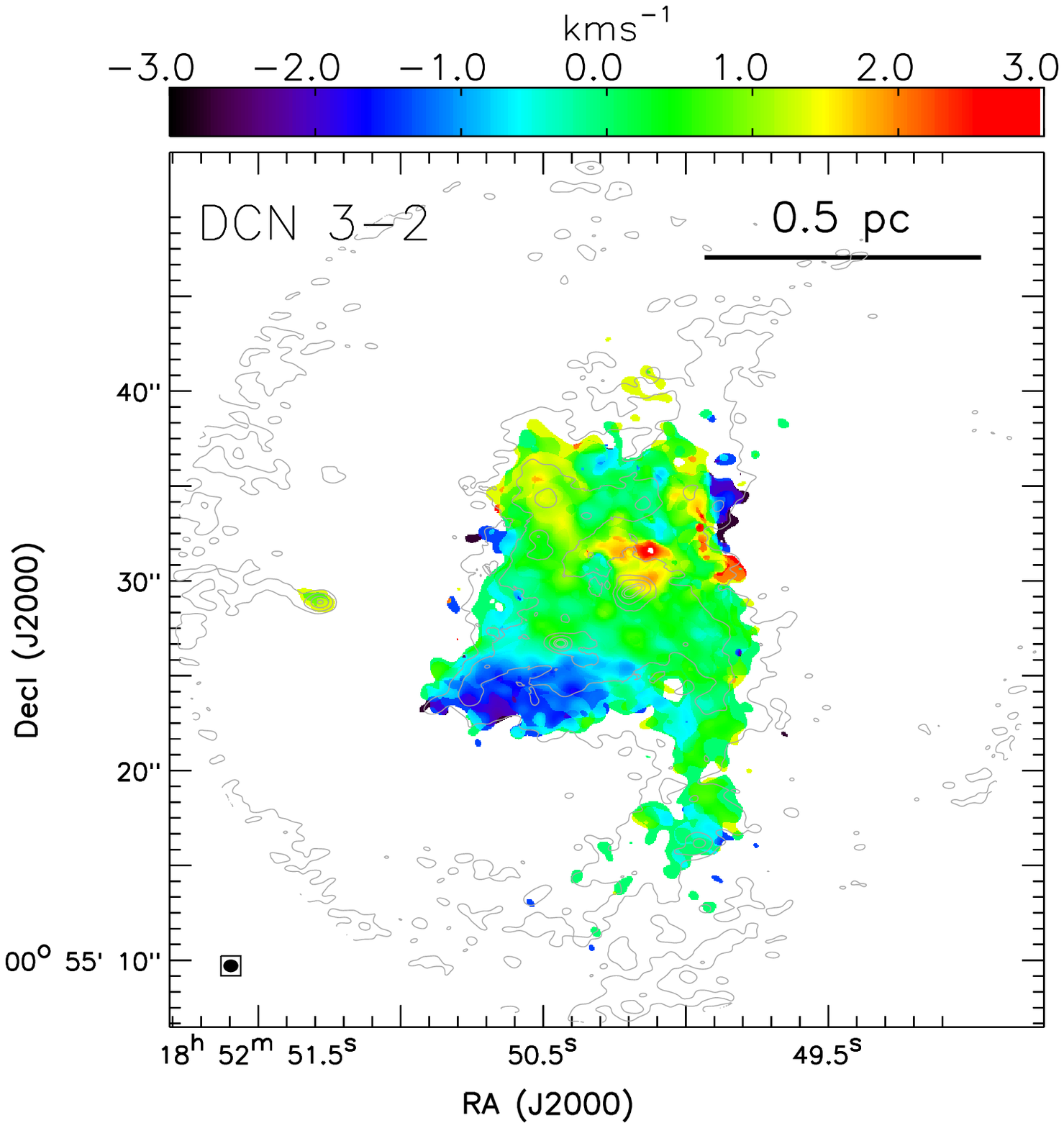} & \includegraphics[width=6.4cm]{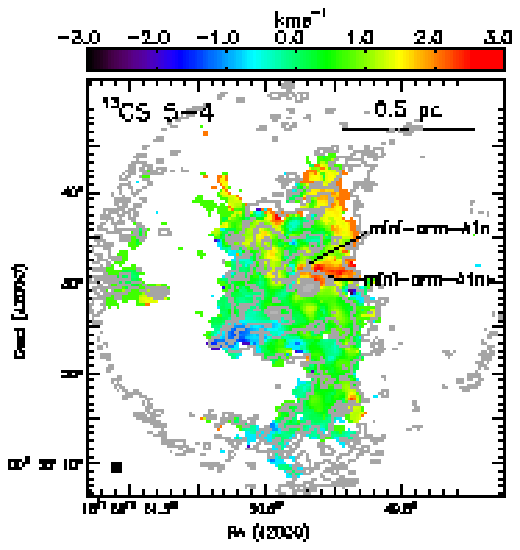} & \includegraphics[width=6.4cm]{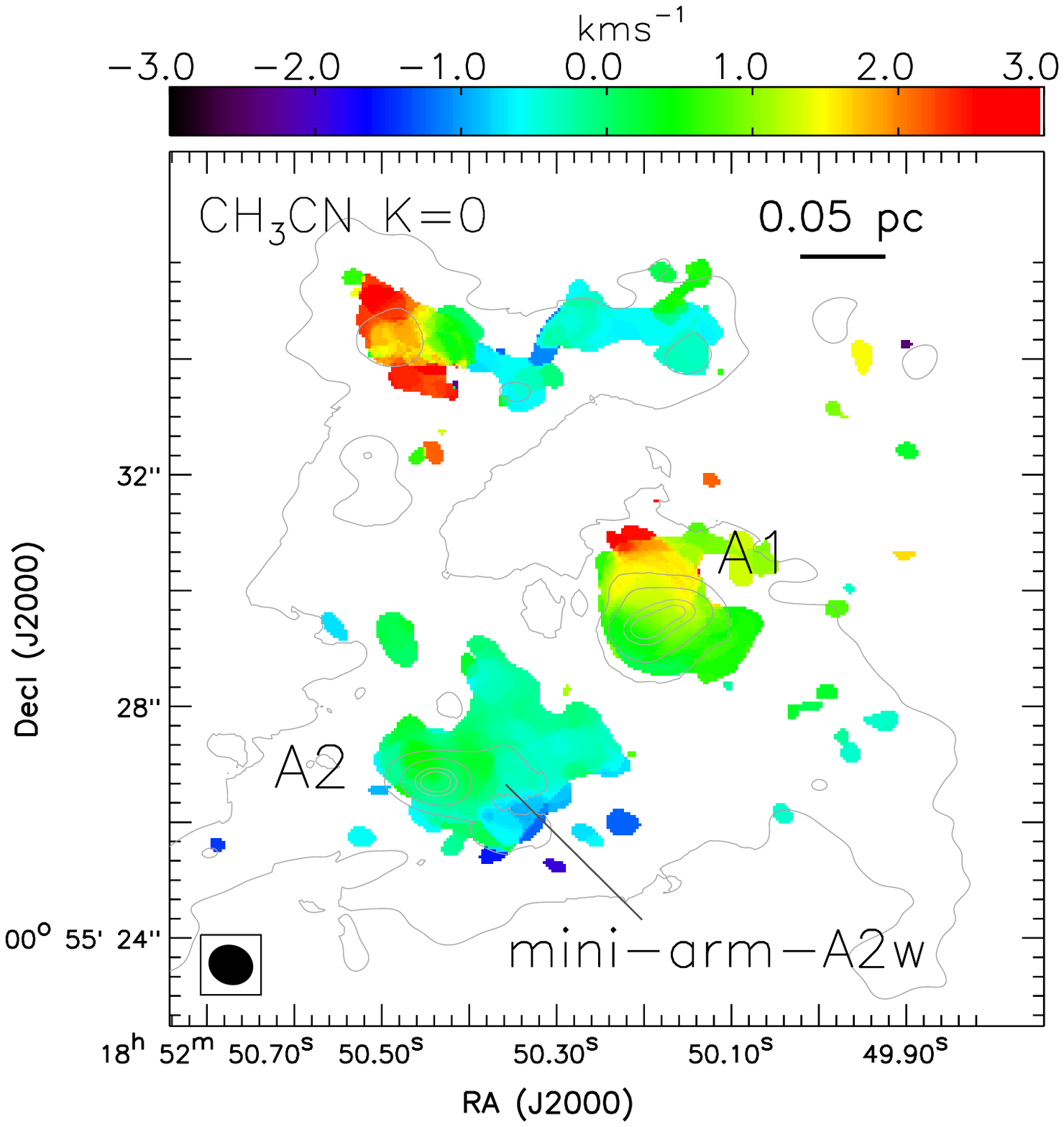}\\ 
\includegraphics[width=6.4cm]{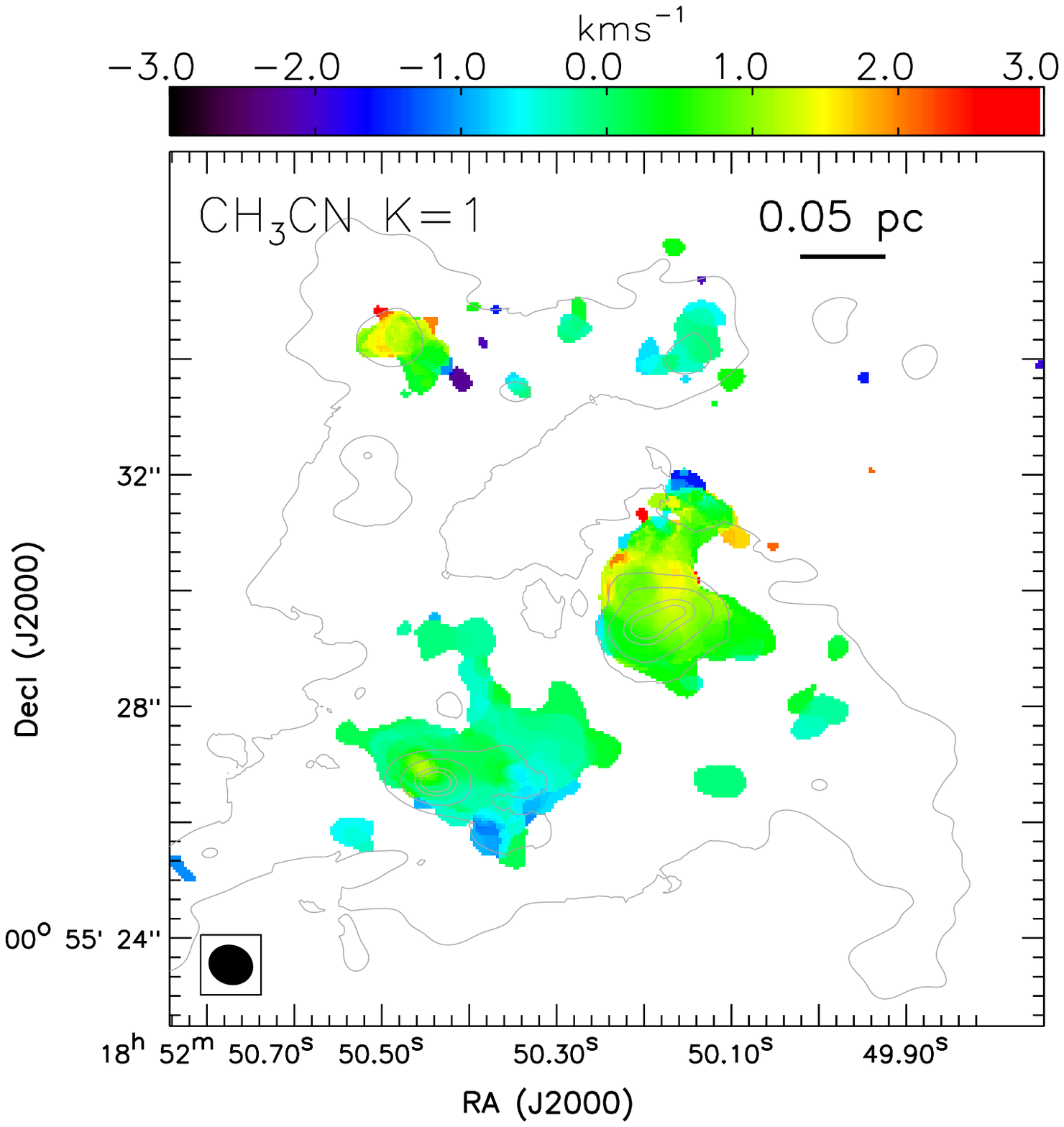} & \includegraphics[width=6.4cm]{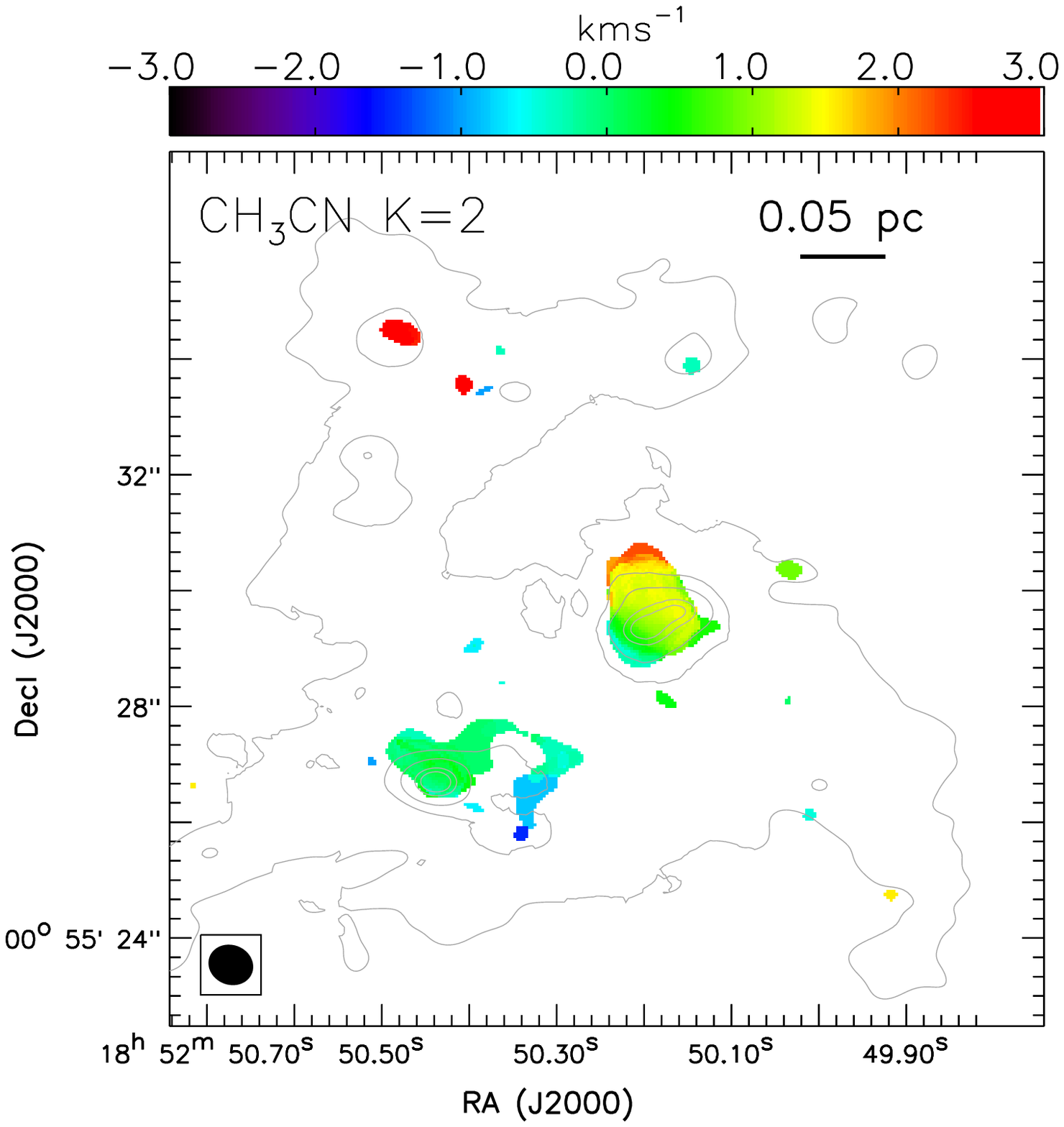} & \includegraphics[width=6.4cm]{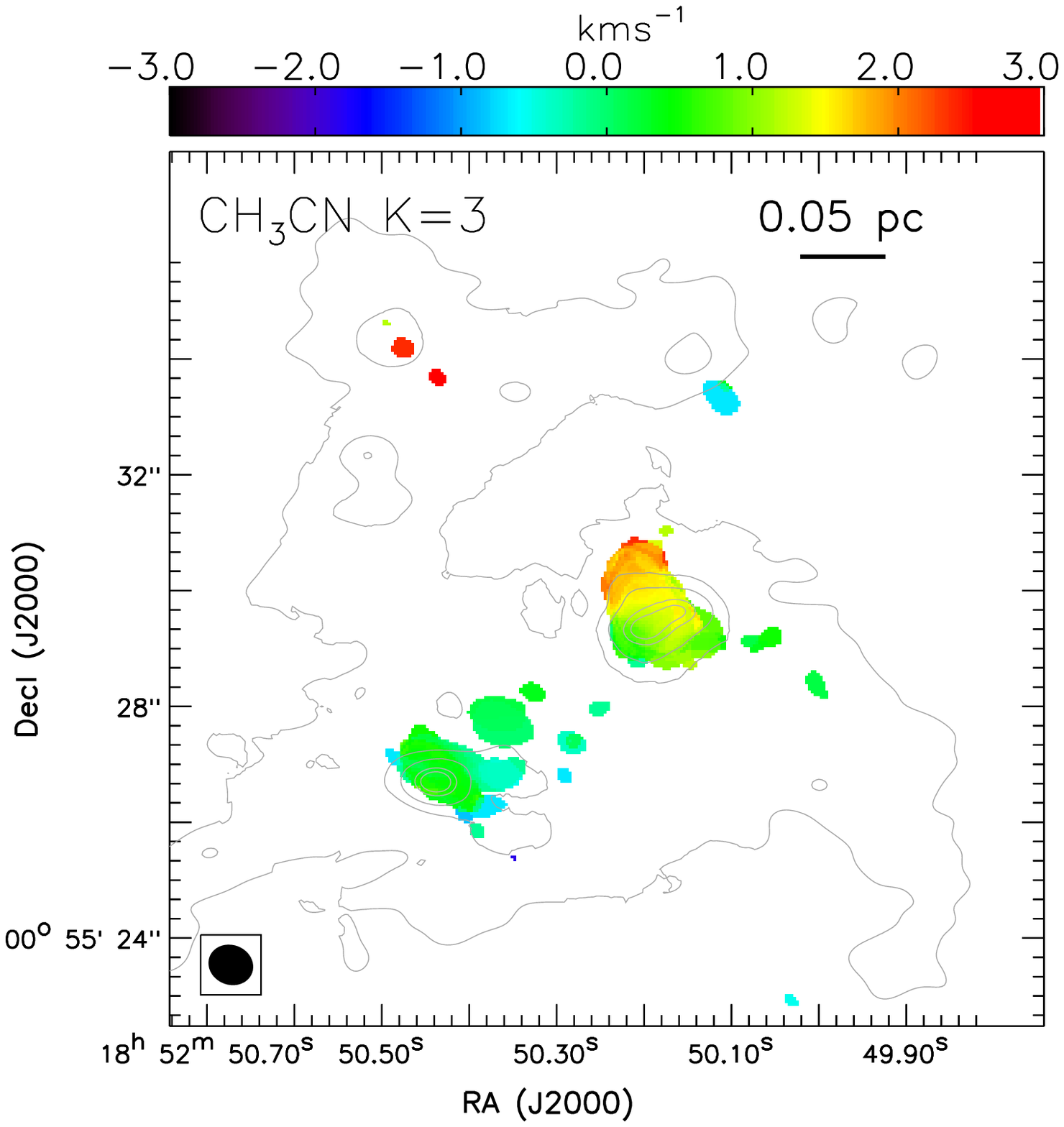}\\ 
\end{tabular}
\caption{\footnotesize{The intensity weighted average velocity maps of the selected molecular lines (color), overlaid on the 1.3 mm dust continuum image (contours). Contours are the same as those in Figure \ref{fig_moment0}. The velocities are relative to $v_{lsr}$=107.6 km\,s$^{-1}$. We note that the blueshifted CH$_{3}$CN K=0 component may be blended with the redshifted K=1 component (see Figure \ref{fig_spectra}).
}} 
\vspace{0.5cm}
\label{fig_moment1}
\end{figure*}

\subsection{Continuum emission} \label{subsec:continuum}
%
%
%
The gas mass is usually estimated with dust thermal continuum emission.
While the continuum emission at the 0.35 mm wavelength is dominated by dust, the continuum emission at 1.3 mm can have contributions from both the dust thermal emission and the free-free emission from the ionized gas.
To separate these two components, we first estimate the free-free emission based on the observations of the hydrogen recombination line H30$\alpha$ line (see Section \ref{subsec:line}).
Assuming that the free-free continuum emission at 1.3 mm is optically thin, the peak hydrogen radio recombination line to free-free continuum flux ratio ($T_{L}$/$T_{C}$) is given by 
\[ 
\sim7.0\times10^{3}(\frac{\delta v}{\mbox{km\,s}^{-1}})^{-1}(\frac{\nu}{\mbox{GHz}})^{1.1}(\frac{T_{e}}{\mbox{K}})^{-1.15}[1+\frac{N(\mbox{He}^{+})}{N(\mbox{H}^{+})}]^{-1}
\]
, where $\delta v$ is the FWHM of the hydrogen radio recombination line, $\nu$ is the observing frequency, and $T_{e}$ is the electron temperature.
We adopt the nominal value of $T_{e}$$\sim$8000 K for the electron temperature, and estimate $T_{L}$ and $\delta v$ based on the observed peak brightness and the second moment in the H30$\alpha$ line image cube.
The assumed $T_{e}$ is an upper limit, since the continuum is over subtracted on small scale when using higher values for $T_{e}$.
The resultant 1.3 mm free-free continuum emission model has a peak flux density of 16.7 mJy\,beam$^{-1}$.
This free-free emission model is then subtracted from the 1.3 mm continuum image to yield the 1.3 mm dust continuum emission image.
We smoothed the 1.3 mm continuum image to the angular resolution of the H30$\alpha$ line image before subtracting the free-free emission model.

Figure \ref{fig_continuum1} and \ref{fig_continuum2} shows the high resolution 0.35 mm continuum image and the 1.3 mm dust continuum image. 
Both continuum images trace the overall gas column density, but the later is optically thinner (i.e. brightness temperature $T_{B}$$\ll$$T_{\mbox{\scriptsize{dust}}}$) thus can penetrate inside the densest regions.
The 0.35 mm continuum image peaks at the location of the $\sim$10$^{4}$ $M_{\odot}$ massive molecular gas clump G33.92+0.11 A (Watt \& Mundy 1999; Liu et al. 2012b).
The peak 0.35 mm flux density is $\sim$73 Jy\,beam$^{-1}$ ($\sim$1.3 K, but can be beam diluted).
The 0.35 mm image resolves the massive companion G33.92+0.11 B northeast of G33.92+0.11 A.
The dominant large scale features are the projected $\sim$5 pc scale elongated structures, connecting from northeast and southwest  (G33-north and G33-south, hereafter) to G33.92+0.11 A and B.
There might be gas filaments connecting G33.92+0.1 A from the west, however, the individual components are not as clearly separated because of confusion with foreground/background gas structures.

\begin{figure*}
\hspace{-0.8cm}
\begin{tabular}{ p{5.9cm} p{5.9cm} p{5.9cm} }
\includegraphics[width=6.4cm]{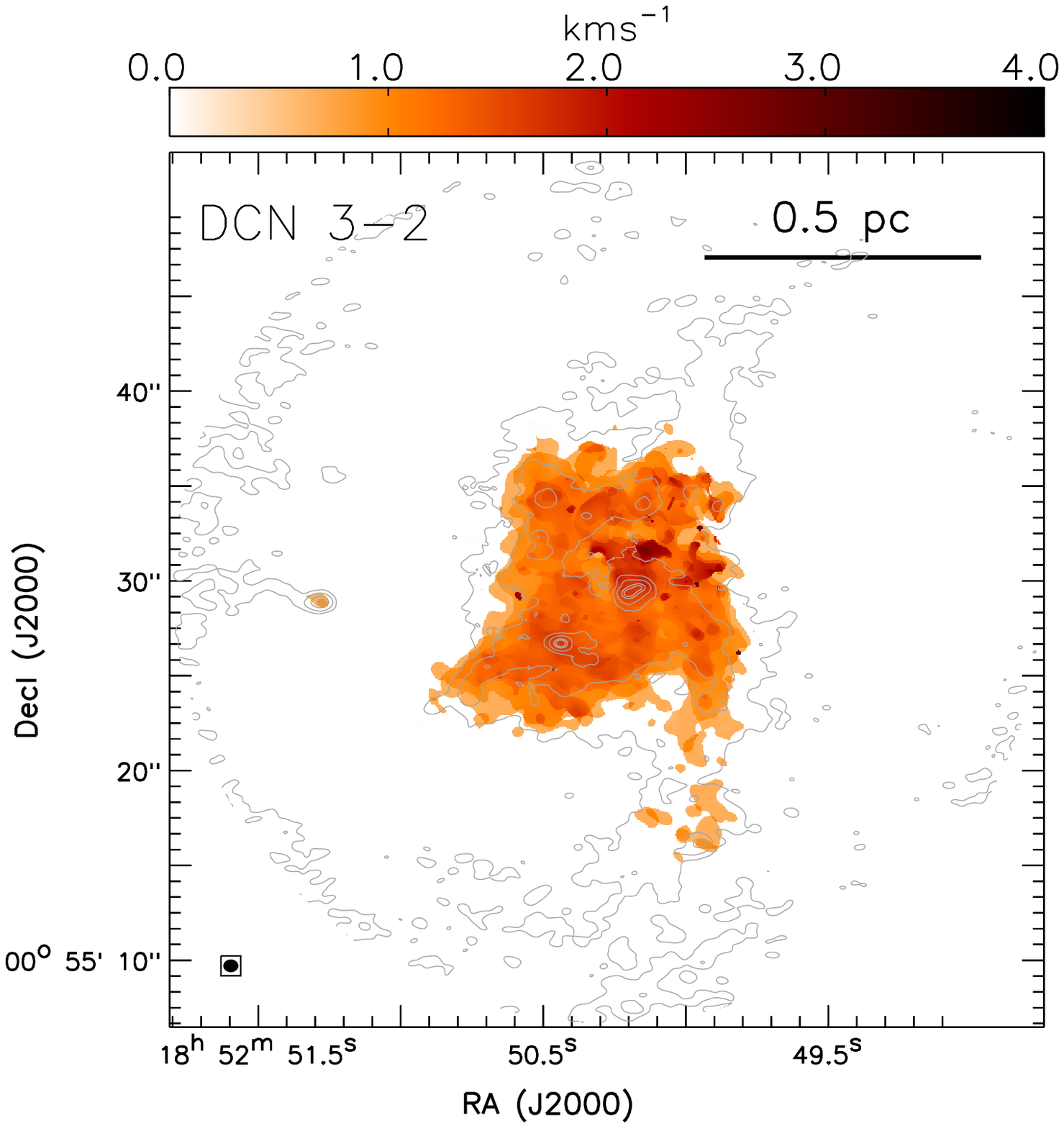} & \includegraphics[width=6.4cm]{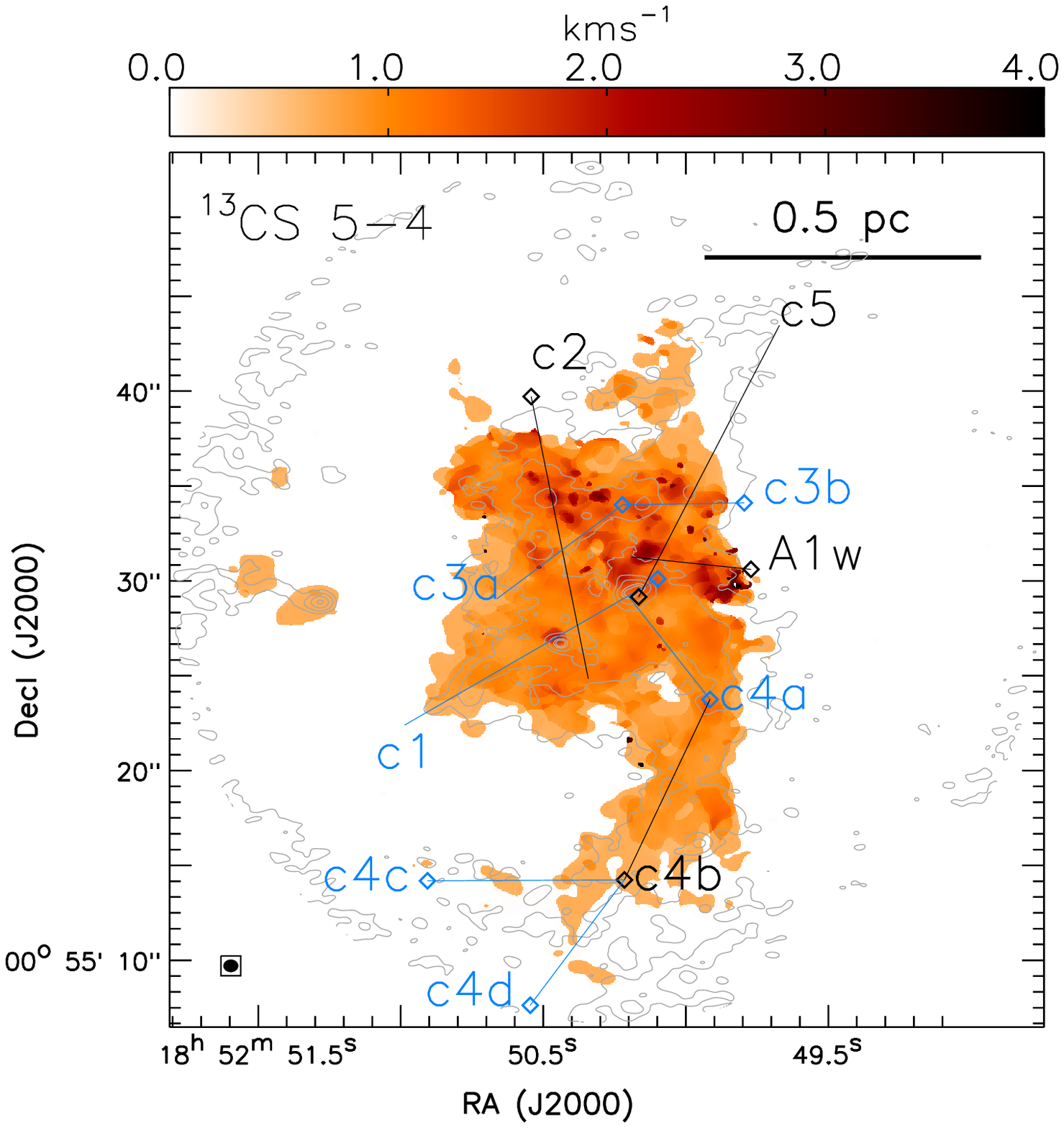} & \includegraphics[width=6.4cm]{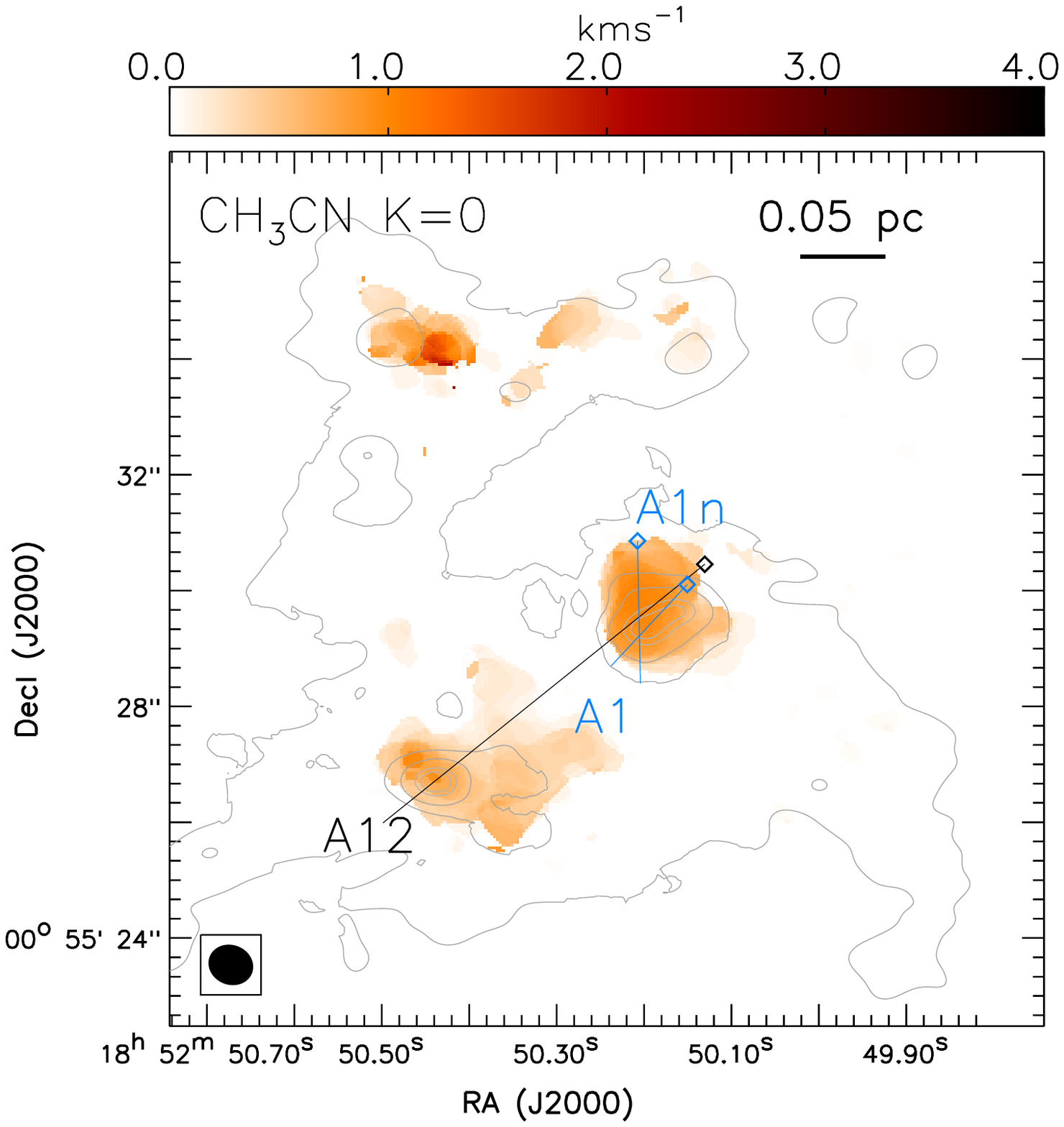}\\ 
\includegraphics[width=6.4cm]{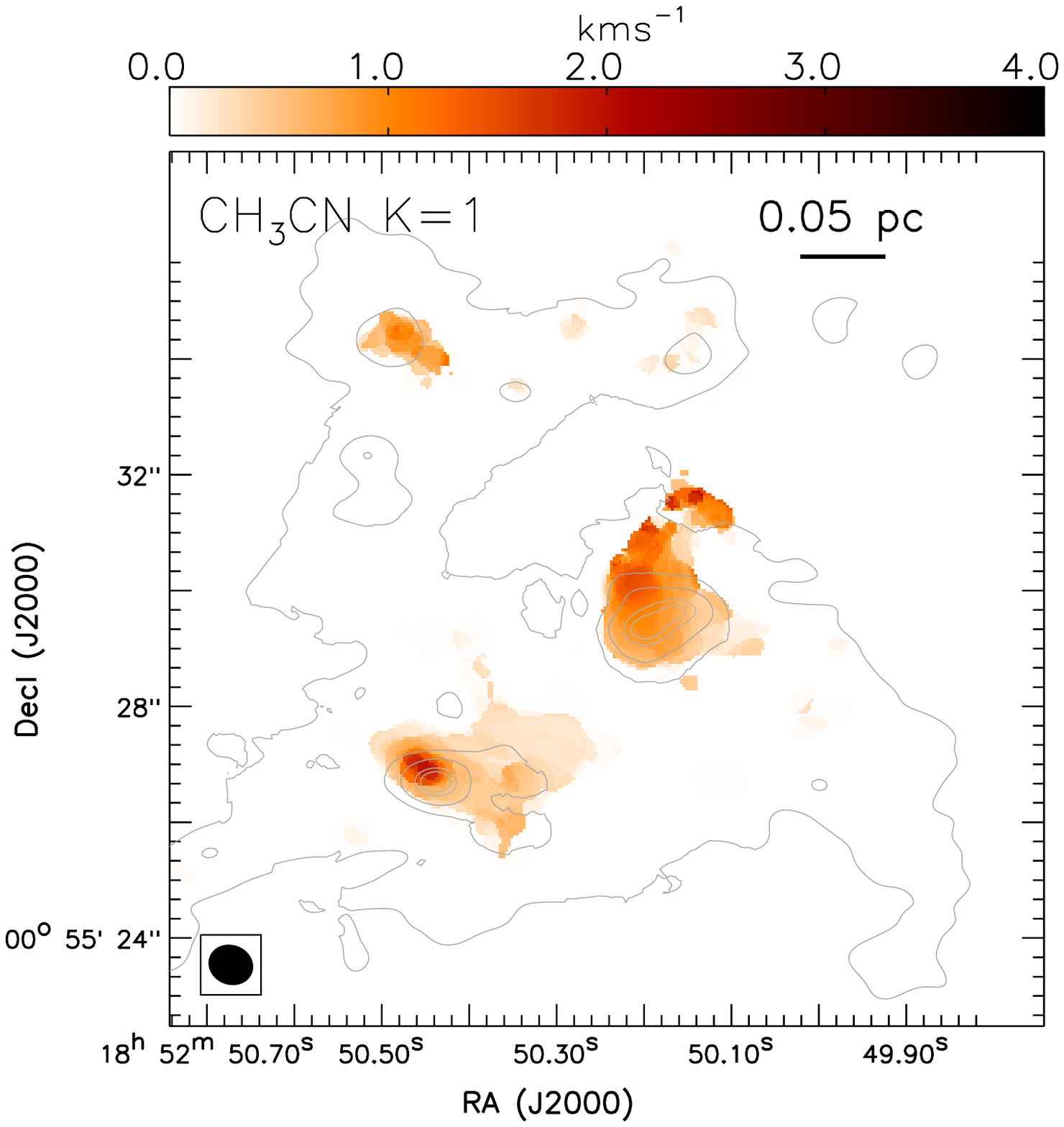} & \includegraphics[width=6.4cm]{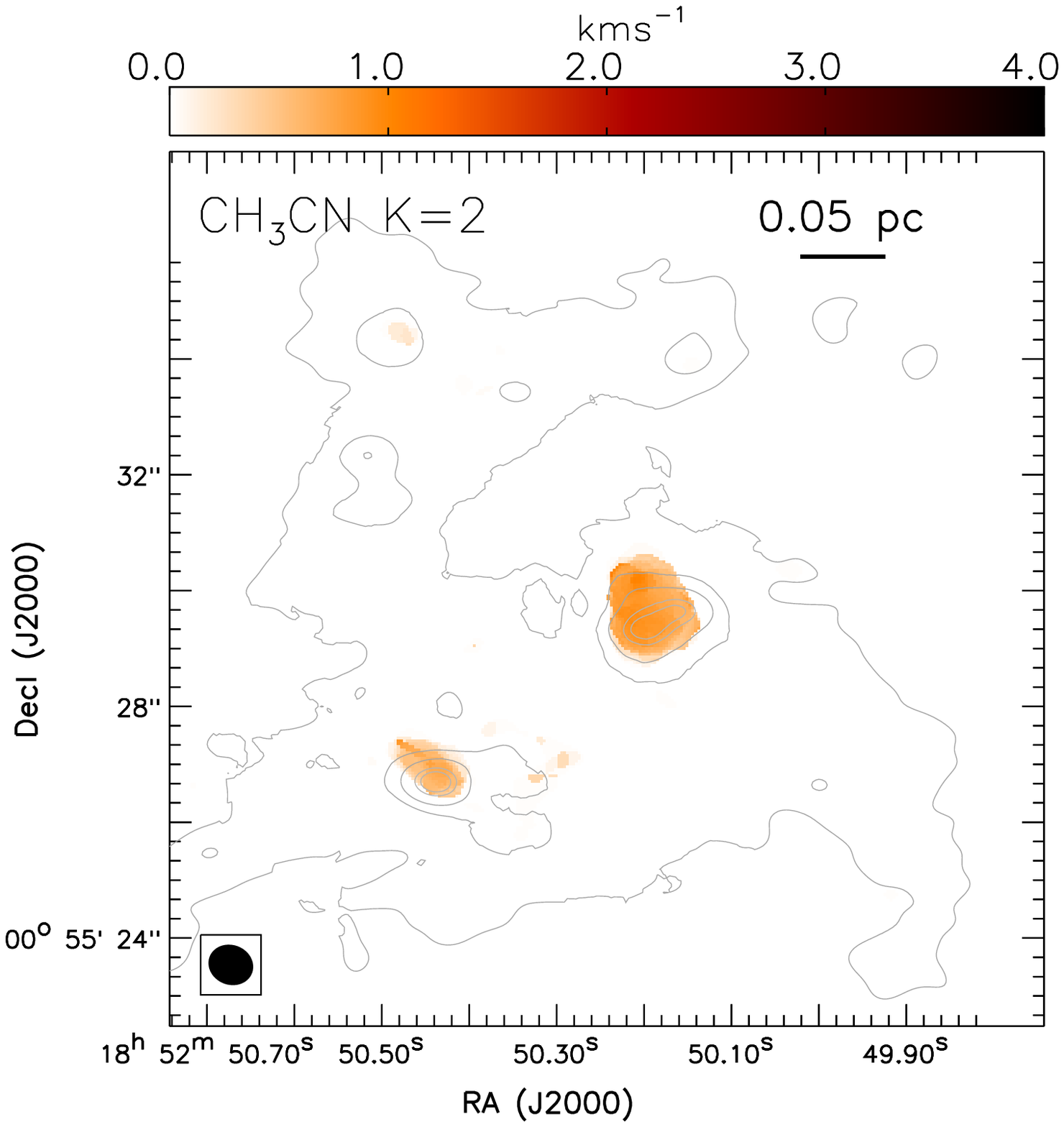} & \includegraphics[width=6.4cm]{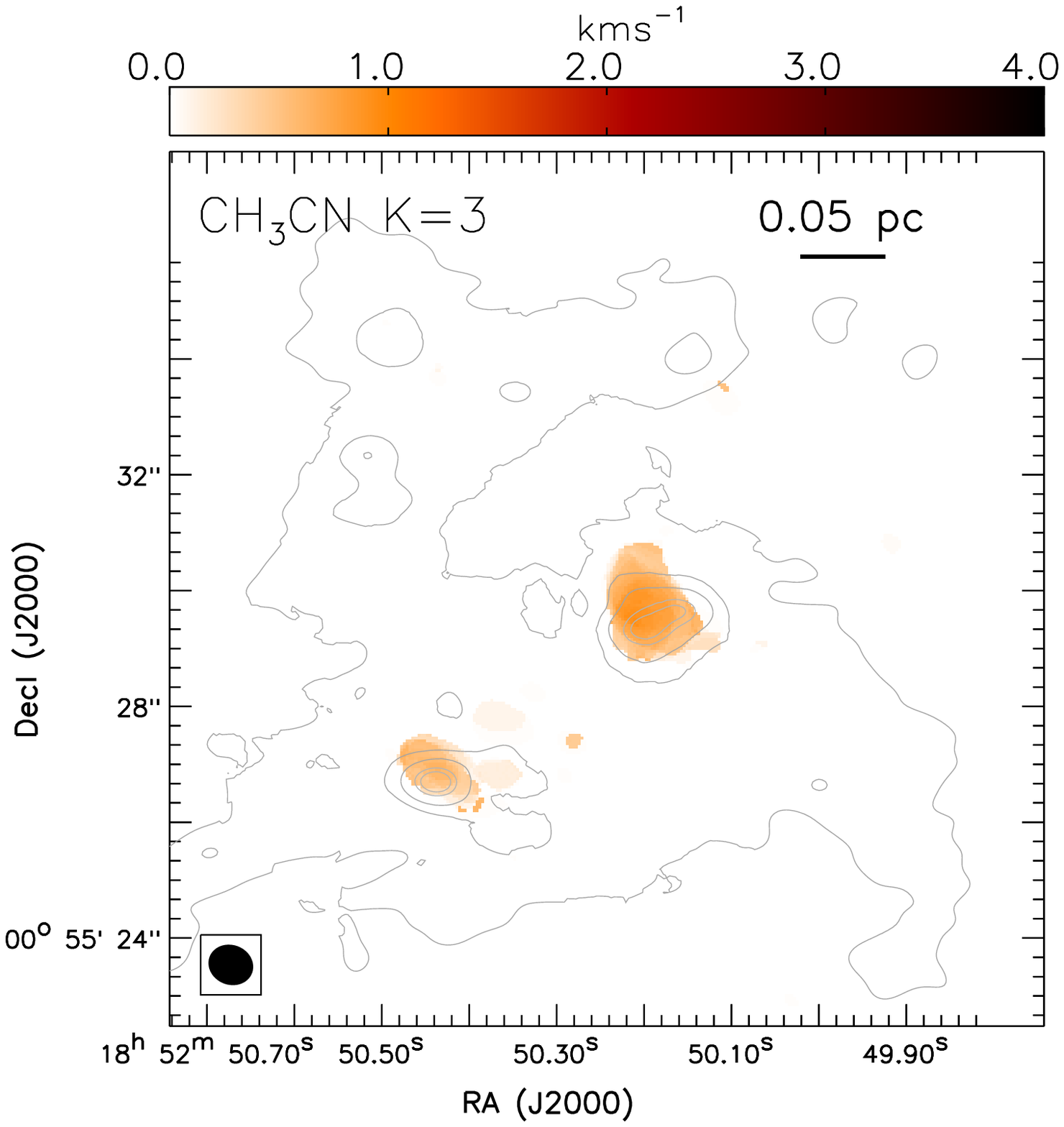}\\ \end{tabular}
\caption{\footnotesize{The intensity weighted velocity dispersion maps of the selected molecular lines (color), overlaid on the 1.3 mm dust continuum image (contours). The velocity dispersion can be converted to full width of half maximum (FWHM) by multiplying the values by a factor of $\sim$2.35. Contours are the same as those in Figure \ref{fig_moment0}. 
We note that the blending of the CH$_{3}$CN K=0, 1 components may artificially enhance the observed velocity dispersion in some regions.
The position-velocity slices used for kinematic analysis are labelled with black and blue lines in the top middle and top right panels.
The generated position-velocity diagrams will be presented in Figure \ref{fig_pvlarge}, \ref{fig_pvmedium}, and \ref{fig_pvsmall}. The colors of these labels are only for avoiding confusion. 
}} 
\vspace{0.5cm}
\label{fig_moment2}
\end{figure*}

\begin{figure*}
\hspace{-0.2cm}
\begin{tabular}{ p{17cm} }
\rotatebox{-90}{
\includegraphics[width=11cm]{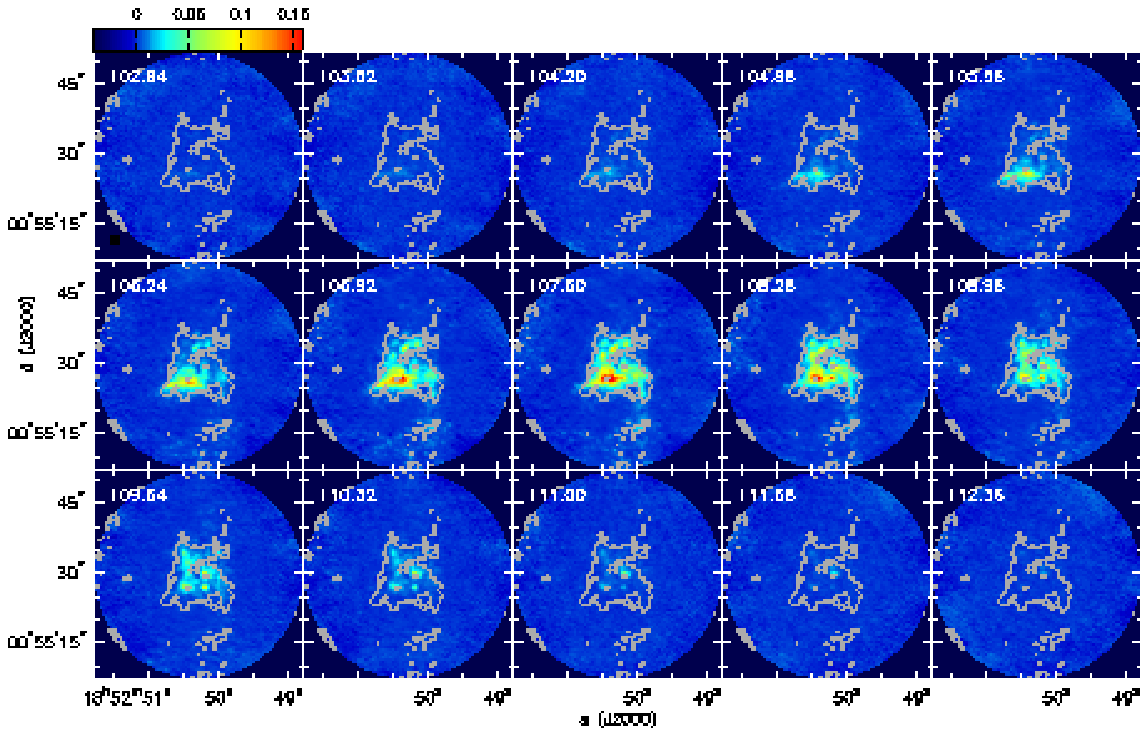} } \\
\end{tabular}
\caption{\footnotesize{The velocity channel maps of the DCN 3-2 line (color), overlaid on the 1.3 mm dust continuum image (contour). Contours are 1.5 mJy\,beam$^{-1}$$\times$[1, 16].
}} 
\vspace{0.3cm}
\label{fig_channeldcn}
\end{figure*}

The 0.6$''$ resolution 1.3 mm dust continuum image (Figure \ref{fig_continuum2}) resolved abundant structures within the inner parsec scale radius.
The most prominent ones are the two 100-300 $M_{\odot}$ massive cores A1 and A2 located at the center, and the five molecular arms roughly following the paths of A1$\rightarrow$A2$\rightarrow$A3 (arm-c1), mini-arm-A2w$\rightarrow$A4$\rightarrow$A5 (arm-c2; see Figure \ref{fig_moment0}, \ref{fig_moment1}), A4$\rightarrow$A6$\rightarrow$A11$\rightarrow$A12$\rightarrow$A13 (arm-c3), A1$\rightarrow$A9 (arm-c4), and A1$\rightarrow$A12 (arm-c5).
The nomenclature here follows Watt \& Mundy (1999) and Liu et al. (2012b), if the referred structures have been detected.
The core A1 is harboring OB stars, which create an UC H\textsc{ii} region.
We refer to Liu et al. (2012b) for a preliminary study of the gas mass in the $<$0.1 pc scale structures.
The molecular arm-c4 appears spiraling out and connected with the parsec scale molecular arm-S1 and S2.
The molecular arm-c2 may be connected with the previously observed molecular arm-N in the north (Liu et al. 2012b), which cannot be imaged well with our ALMA 12m Array data due to the smaller primary beam size.
Following the convention in Liu et al. (2012b), we refer to the OB stars that are embedded in either core A1 or core A2 as the \textit{central OB cluster} hereafter. 
For the rest of localized intermediate- or high-mass (proto)stars and their parent molecular cores in G33.92+0.11 A, we refer to them as the \textit{satellite} high-mass stars and satellite cores.
The previously identified core A4 is now resolved into two components A4n and A4s. 
In addition to the aforementioned gas structures, we also resolved the $\sim$0.1 pc scale gas mini-arms connecting with the cores.
The gas mini-arms connecting A2 from its west (mini-arm-A2w) have been resolved in the previous SMA observations (Liu et al. 2012b).
There are mini-arms connecting A1 from its north and northwest (mini-arm-A1n, mini-arm-A1nw), which can only be marginally seen in Figure \ref{fig_continuum1} owing to blending with other structures. 
We will address more about these gas mini-arms based on the spectral line results (Section \ref{subsec:line}).

The high dynamic range of the ALMA observations confirms that the two highest mass cores A1 and A2 in the center are not round.
The core A1 appears slightly elongated in the southeast-northwest direction. 
This may imply that it is an inclined flattened structure, or there possesses eccentric gas orbits.
But we cannot rule out the possibility that there is internal fragmentation in core A1, or that there are more than two cores blended in the projected line-of-sight.
We cannot resolve the internal structure of all detected cores.
Some, or all of the individual cores may eventually form a cluster of stars. 
The complex morphology resolved by ALMA has suggested that the accretion flows in the massive molecular clump is highly dynamical/chaotic. 
In fact, the resolved arm-like features may resemble the scaled-down version of eccentric gas arms/arcs connected with the 2-4 pc scale Galactic circumnuclear disk/ring (Liu et al. 2012d).


\begin{figure*}
\hspace{-0.2cm}
\begin{tabular}{ p{17cm} }
\rotatebox{-90}{
\includegraphics[width=11cm]{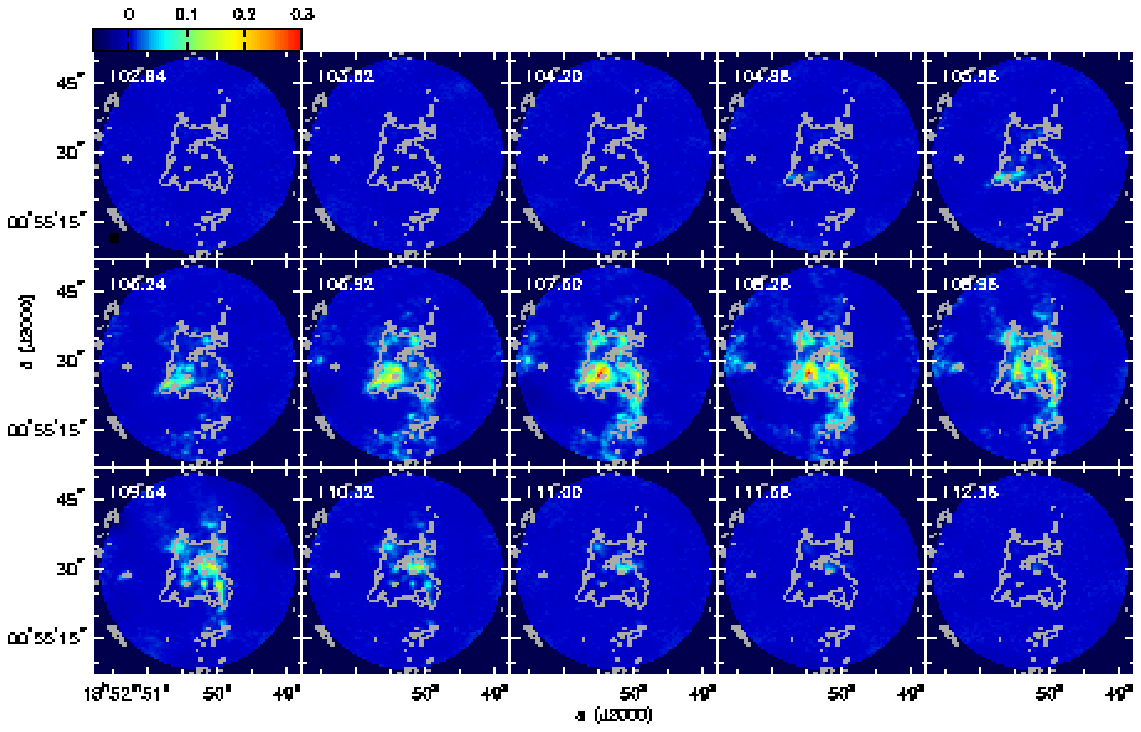} } \\
\end{tabular}
\caption{\footnotesize{The velocity channel maps of the $^{13}$CS 5-4 line (color), overlaid on the 1.3 mm dust continuum image (contour). Contours are 1.5 mJy\,beam$^{-1}$$\times$[1, 16].
}} 
\vspace{0.3cm}
\label{fig_channel13cs}
\end{figure*}

\subsection{Spectral lines} \label{subsec:line}
\begin{figure*}
\hspace{-0.2cm}
\begin{tabular}{ p{17cm} }
\rotatebox{-90}{
\includegraphics[width=17cm]{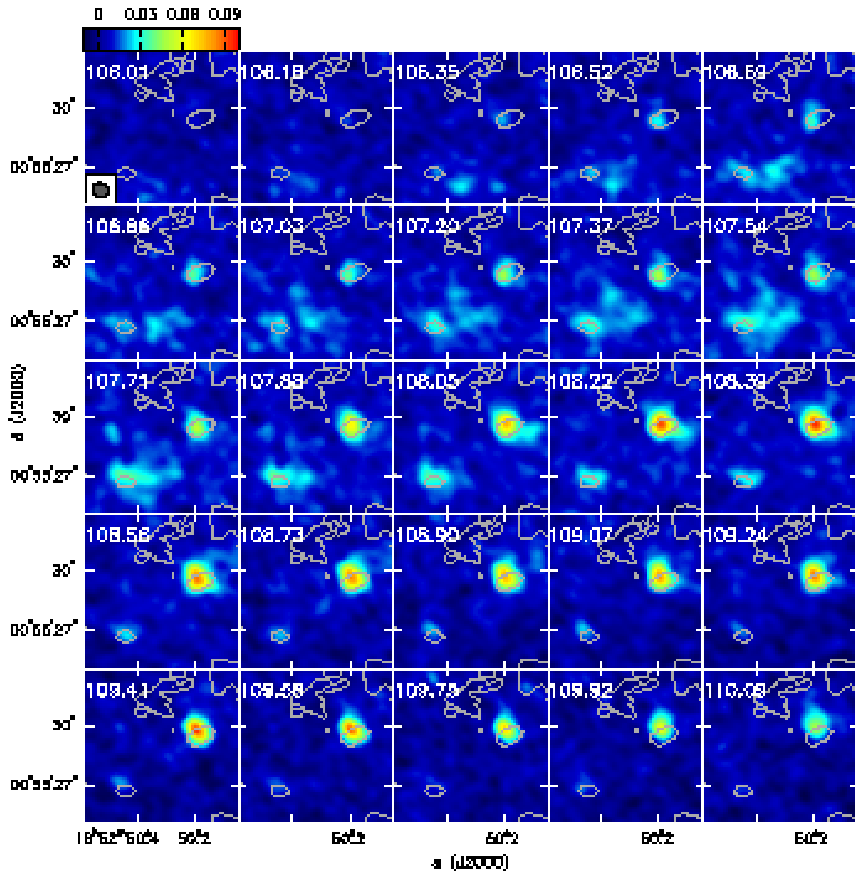} } \\
\end{tabular}
\caption{\footnotesize{The velocity channel maps of the CH$_{3}$CN J=12-11, K=0 line (color), overlaid with the 1.3 m dust continuum image (contour). Contours are 1.5 mJy\,beam$^{-1}$$\times$[1, 16].
}} 
\vspace{1.2cm}
\label{fig_channelch3cnk0}
\end{figure*}

We present the full spectral window spectra generated from the central 10$''$ field of our ALMA 12m+ACA observations, in Figure \ref{fig_spectra}.
The identified line species are listed in Table \ref{tab-lines}.
In this work, we focus on the discussion of the CH$_{3}$CN J=12-11 K-ladders, the $^{13}$CS 5-4 line, and the DCN 3-2 line.
The velocity integrated intensity maps (i.e. moment 0) of these lines are shown in Figure \ref{fig_moment0}.
Figure \ref{fig_rgb} provides the side-by-side comparison of the blowed-up line images and the 1.3 mm dust continuum image in the central parsec scale area.
The intensity weighted velocity maps (i.e. moment 1), and the intensity weighted velocity dispersion maps (i.e. moment 2) of these lines are presented in Figures \ref{fig_moment1}--\ref{fig_moment2}.
We present the velocity channel maps of the DCN 3-2 line, and of the $^{13}$CS 5-4 line in Figure \ref{fig_channeldcn} and \ref{fig_channel13cs}.
The position-velocity (PV) diagrams of these two lines, generated from the selected slices (see Figure \ref{fig_moment2}), are shown in  Figure \ref{fig_pvlarge}.
We present the velocity channel maps of the lowest and the highest transitions of the detected CH$_{3}$CN J=12-11 K-ladders in Figures \ref{fig_channelch3cnk0} and \ref{fig_channelch3cnk3}.
The PV diagrams for these CH$_{3}$CN transitions are provided in Figure \ref{fig_pvmedium} and Figure \ref{fig_pvsmall}.
We introduce the observed intensity distribution and velocity profiles separately in Section \ref{subsubsec:line_dist} and \ref{subsubsec:line_vel}.

\subsubsection{Distribution}\label{subsubsec:line_dist}
The spatial distribution of the dense gas tracer DCN 3-2 is in excellent agreement with the distribution of the 1.3 mm dust emission structures (Figure \ref{fig_moment0}, \ref{fig_channeldcn}). 
Strong DCN 3-2 emission is seen on the previously identified molecular clump C (Liu et al. 2012b), all aforementioned molecular arms, and shows several emission peaks coincide with the molecular cores A1--A6 and A11.
The $^{13}$CS 5-4 line emission is also detected towards all these structures. 
However, the integrated $^{13}$CS 5-4 map does not show enhanced emission from arm-c2 and arm-c3.
From Figure \ref{fig_moment0}, the most prominent $^{13}$CS 5-4 emission structures are the mini-arm-A1n, mini-arm-A1nw, arm-c5, mini-arm-A2w, and the north part of arm-c1.
The former three (mini-)arms engulf the converging point between arm-c4 and core A1, and the later two surround the core A2.
The $^{13}$CS 5-4 enhanced mini-arms surrounding core A1 are all redshifted relative to the systemic velocity $v_{lsr}^{\mbox{\scriptsize{sys}}}$ of 107.6 km\,s$^{-1}$ (more about the systemic velocity see Liu et al. 2012b and references therein).
They start to emerge from the ambient gas structures in the velocity channel of 108.28 km\,s$^{-1}$ (Figure \ref{fig_channel13cs}), and becomes the brightest structures in the velocity channels redder than 110.32 km\,s$^{-1}$.
The redshifted mini-arm-A1nw can directly be seen in the average velocity map (Figure \ref{fig_moment1}), which artificially present a high velocity dispersion because of blending with the ambient gas structures along the line-of-sight (Figure \ref{fig_moment2}). 
Toward the same area, we also found high NH$_{3}$ rotational temperature (Liu et al. 2012b).
The mini-arm-A1n and mini-arm-A1nw are also weakly detected in the velocity maps of DCN 3-2 (Figure \ref{fig_moment1}, Figure \ref{fig_channeldcn}), which may indicate that these warm mini-arms are the result of the quickly dynamically processed gas streams. 
The redshifted motions of the mini-arm-A1nw can only be seen in its west end in the average velocity map of DCN 3-2 (Figure \ref{fig_moment1}, top left), because of large confusion with the bulk motions of the ambient gas.
The projected geometry of core A1 together with arm-c4, mini-arm-A1w and mini-arm-A1nw highlighted by $^{13}$CS 5-4, resembles the scaled-down geometry of G33.92+0.11 A connected with the arm-N, S1, and S2 seen in the previous VLA NH$_{3}$ images and the SMA images of the CO 2-1 isotopologues (Liu et al. 2012b).
The $^{13}$CS 5-4 line does not directly trace core A5, but some extended gas structures adjacent to it.
This may suggest that core A5 is surrounded by extended shocked gas.

We only detected the K=0, 1, 2, 3 components of the CH$_{3}$CN J=12-11 lines with upper level energies of 69 K, 76 K, 97 K, and 133 K, respectively.
These lines are enhanced towards core A1, A2, and A5, but an extended and fainter component is seen in the entire $\sim$0.6 pc area in G33.92+0.11 A.
In particular, all K components of CH$_{3}$CN J=12-11 trace emission in mini-arm-A2w with an extension to its northwest.
The color composited images in Figure \ref{fig_rgb} clearly demonstrate the similarity and the difference between the presented molecular gas tracers.

\subsubsection{Velocity Profile}\label{subsubsec:line_vel}

\begin{figure*}
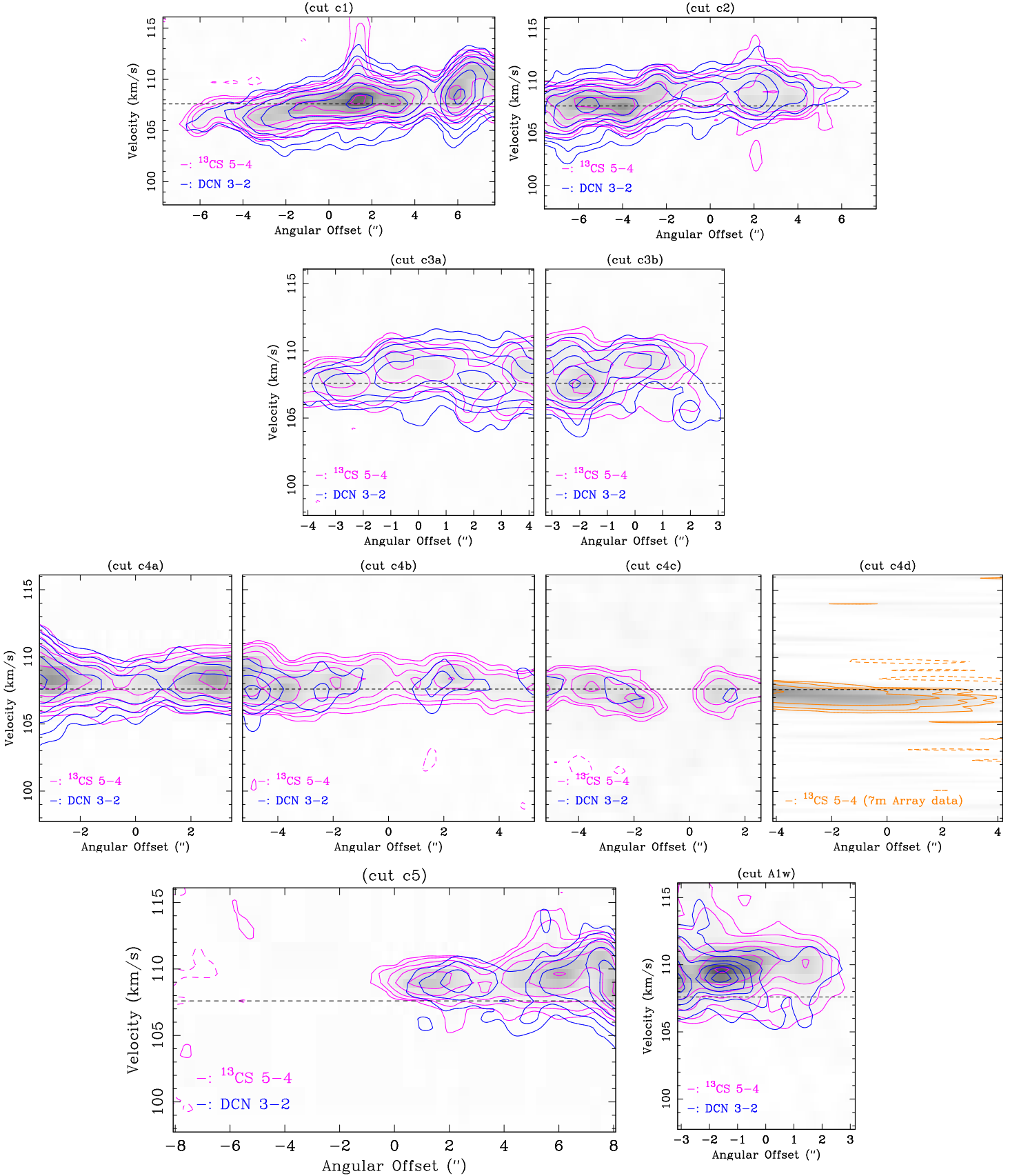

\hspace{1.5cm}
\begin{tabular}{ p{7cm} p{7cm} }
\rotatebox{-90}{\includegraphics[scale=0.3]{pv_cutc1.ps}} & \rotatebox{-90}{\includegraphics[scale=0.3]{pv_cutc2.ps}} \\ 
\end{tabular}

\hspace{4.2cm}
\begin{tabular}{ p{5cm} p{3cm} }
\rotatebox{-90}{\includegraphics[scale=0.3]{pv_cut3a.ps}} & \rotatebox{-90}{\includegraphics[scale=0.3]{pv_cut3b.ps}} \\ 
\end{tabular}

\hspace{-0.85cm}
\begin{tabular}{ p{4.25cm} p{5.5cm} p{3.95cm} p{4cm}}
\rotatebox{-90}{\includegraphics[scale=0.3]{pv_cut4a.ps}} & \rotatebox{-90}{\includegraphics[scale=0.3]{pv_cut4b.ps}} & \rotatebox{-90}{\includegraphics[scale=0.3]{pv_cut4c.ps}} & \rotatebox{-90}{\includegraphics[scale=0.3]{pv_cut4d.ps}} \\ 
\end{tabular}

\hspace{1.5cm}
\begin{tabular}{ p{9.6cm} p{9cm} }
\rotatebox{-90}{\includegraphics[scale=0.4]{pv_cut5.ps}} & \rotatebox{-90}{\includegraphics[scale=0.3]{pv_cuta1w.ps}} \\ 
\end{tabular}

\caption{\footnotesize{ 
Position-Velocity (PV) diagrams of the $^{13}$CS 5-4 (grayscale and magenta contours) and the DCN 3-2 lines (blue contours). Contours are 4.8 mJy\,beam$^{-1}$(3$\sigma$)$\times$[-1, 1, 2, 4, 8, 16, 32, 64], except for the DCN 3-2 contours in the `cut A1w' panel, which are 4.8 mJy\,beam$^{-1}$(3$\sigma$)$\times$[1, 2, 3, 4, 5, 6]. Dashed lines label the $v_{lsr}$=107.6 km\,s$^{-1}$. The PV cuts are drawn in Figure \ref{fig_moment2}, with diamonds indicating the end points of the cuts (i.e. positive angular offset). Zero position offset is defined at the center of the presented PV cuts. Orange contours (0.1 Jy\,beam$^{-1}$(4$\sigma$)$\times$[-1, 1, 2, 4, 8]) present the PV diagram of the `cutc4d' using only the ALMA 7m Array data. 
}} 
\vspace{0.5cm}
\label{fig_pvlarge}
\end{figure*}

DCN trace dense gas, where the ices have been recently evaporated and injected into the gas phase. 
Since the destruction of deuterated neutral species in warm gas is slow ($\sim$10$^{4}$--10$^{5}$ yrs), DCN represents a fossil record of the deuterium content in the ices when the region was cold (Rodgers \& Millar 1996).
The $^{13}$CS 5-4 line traces large gas volume densities and large column densities, due to its high critical density and low abundance.
Therefore, these two lines can probe the motions in the large scale gas flows, without being seriously confused by localized (proto)stellar feedback.

The velocity maps of the DCN 3-2 line and the $^{13}$CS 5-4 line consistently trace a southeast-northwest velocity gradient in the $\sim$0.6 pc scale massive molecular clump G33.92+0.11A (Figure \ref{fig_moment1}, \ref{fig_channeldcn}, \ref{fig_channel13cs}). 
The parsec scale molecular arm-S1,S2 appear blueshifted in the south and southeast ends, and become redshifted when moving closer to the dense molecular clump.
Spatially, this is the most easily seen in the $^{13}$CS 5-4 velocity channel map.
The gas motion within the molecular arms are presented in the position-velocity (PV) diagrams of the $^{13}$CS 5-4 and the DCN 3-2 lines (Figure \ref{fig_pvlarge}).
We carefully placed the PV cuts, such that we can see the velocity gradient within the targeted molecular arms, without being confused much by the ambient dense gas structures. 
The locations of the presented PV cuts are drawn in the top middle panel of Figure \ref{fig_moment2}.
The PV diagram from the `cut c4d' (Figure \ref{fig_moment2}, \ref{fig_pvlarge}) is generated from the high spectral resolution (0.16 km\,s$^{-1}$) $^{13}$CS 5-4 data taken with the ALMA 7m Array alone ($\theta_{\mbox{\scriptsize{maj}}}$$\times$$\theta_{\mbox{\scriptsize{min}}}$=7$''$.1$\times$4$''$.5, P.A.=89$^{\circ}$), which is more sensitive to the mildly blueshifted motion in the south because of the larger field of view than the ALMA 12m Array observations.

From Figure \ref{fig_pvlarge}, we see that although the DCN 3-2 and the $^{13}$CS 5-4 lines do not trace the same localized structures, the large scale velocity gradients traced by these two lines appear consistent. 
On the small scale, we see that the $^{13}$CS 5-4 line traces slightly more redshifted emission by $\sim$1 km\,s$^{-1}$ in the molecular gas mini-arm-A1nw, as shown by the PV `cut A1w'.

\begin{figure*}
\hspace{-0.2cm}
\begin{tabular}{ p{17cm} }
\rotatebox{-90}{
\includegraphics[width=17cm]{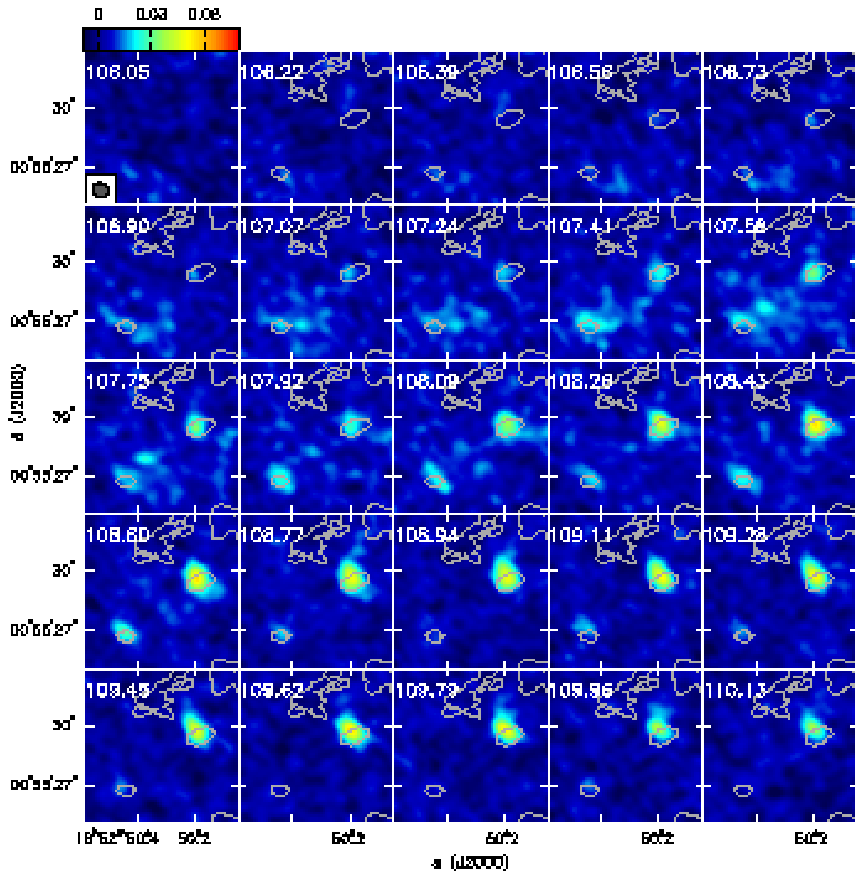} } \\
\end{tabular}
\caption{\footnotesize{The velocity channel maps of the CH$_{3}$CN J=12-11, K=3 line (color), overlaid with the 1.3 m dust continuum image (contour). Contours are 1.5 mJy\,beam$^{-1}$$\times$[1, 16]. 
}} 
\vspace{1.2cm}
\label{fig_channelch3cnk3}
\end{figure*}

From `cut c1', `cut c2', `cut c4a', and `cut c5', we do not resolve the Keplerian-like velocity profile.
Rather than converging towards the cloud systemic velocity $v_{lsr}^{\mbox{\scriptsize{sys}}}$ at outer radii, we see the approximately linear velocity gradients in the molecular arms-c1, c2, and c5,  which show the high velocity offsets at locations further away from the center of G33.92+0.11 A.
As a result, the average velocity map of $^{13}$CS 5-4 (Figure \ref{fig_moment1}, top middle) trace the very blueshifted and redshifted motion in southeast and northwest of G33.92+0.11 A, respectively. 
In the average velocity map of DCN 3-2 (Figure \ref{fig_moment1}, top left), the redshifted motion in the northwest is confused by some diffused blueshifted high velocity features, which have been reported in the previous observations (Liu et al. 2012b).
From `cut c3a' and `cut c3b' (Figure \ref{fig_pvlarge}), we see that the molecular arm-c3 is redshifted relatively to $v_{lsr}^{\mbox{\scriptsize{sys}}}$, and shows increasing velocity towards the northwest.
The higher redshifted feature around the -0.8$''$ offset of `cut c3a' is likely contributed by arm-c2.

From the CH$_{3}$CN velocity maps (Figure \ref{fig_moment1}), we see that the core A1 is in general redshifted relative to core A2. 
Gaussian fits of the CH$_{3}$CN spectral line profiles found the $v_{lsr}$ of cores A1 and A2 to be 108.93$\pm$1.19 km\,s$^{-1}$, and 107.55$\pm$0.83 km\,s$^{-1}$, respectively. 
However, the CH$_{3}$CN spectral profile in core A1 is red-skewed, which may be due to spatial blending with the redshifted motion of mini-arm-A1n, or self-absorption.
The centroid velocity of core A1 determined by averaging the observed minimal and maximal terminal velocities, is $\sim$108.47 km\,s$^{-1}$.
The mini-arm-A2w has a blueshifted velocity of 107.19$\pm$0.83 km\,s$^{-1}$.

\begin{figure}
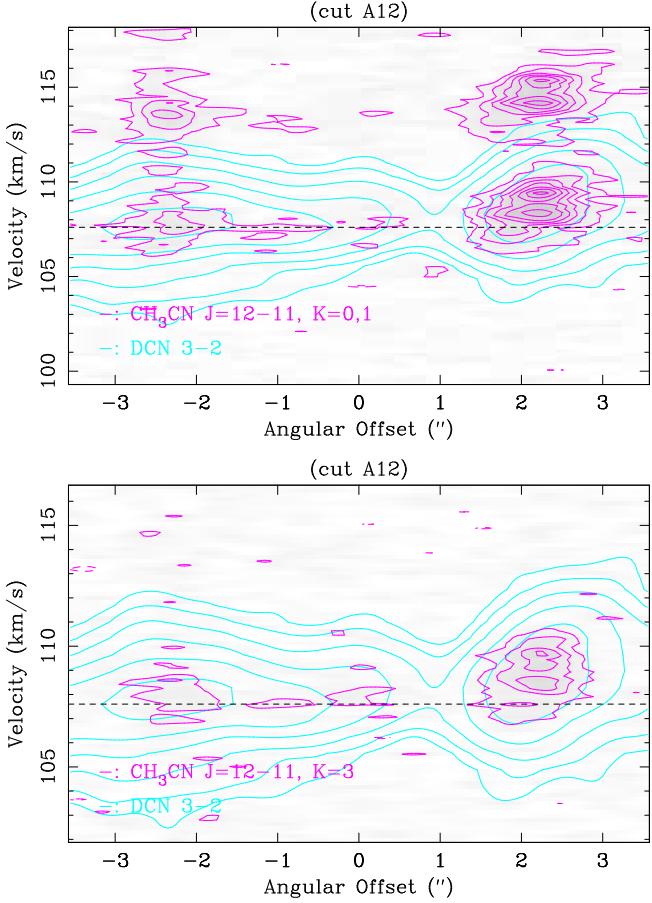


\hspace{-0.4cm}
\begin{tabular}{ p{5cm}  }
\rotatebox{-90}{\includegraphics[scale=0.362]{pv_cuta12_k01.ps}}   \\ 
\end{tabular}

\hspace{-0.4cm}
\begin{tabular}{ p{5cm} }
\rotatebox{-90}{\includegraphics[scale=0.362]{pv_cuta12_k3.ps}}    \\ 
\end{tabular}

\caption{\footnotesize{Position-Velocity (PV) diagram for the CH$_{3}$CN J=12-11, K=0,1,3 lines (magenta contour and grayscale) made within the central 0.3 pc radius in G33.92+0.11 A, overlaid with the PV diagram of the DCN 3-2 line (cyan contour) generated from the same PV cuts. Dashed lines label the $v_{lsr}$=107.6 km\,s$^{-1}$. The PV diagrams of CH$_{3}$CN lines are generated from the Briggs robust=0 weighted image cubes, which has a synthesized beam of $\theta_{\mbox{\scriptsize{maj}}}$$\times$$\theta_{\mbox{\scriptsize{min}}}$=0$''$.73$\times$0$''$.51 (P.A.=77$^{\circ}$). The K=0 and 1 components of CH$_{3}$CN J=12-11 are presented in the same panel, of which the velocity axis is alined with that of the K=0 component. The PV cuts are drawn in Figure \ref{fig_moment2}, with diamonds indicating the end points of the cuts (i.e. positive angular offset). Magenta contours are 10.5 mJy\,beam$^{-1}$(3$\sigma$)$\times$[-1, 1, 2, 3, 4, 5, 6, 7, 8]. Cyan contours are 4.8 mJy\,beam$^{-1}$(3$\sigma$)$\times$[-1, 1, 2, 4, 8, 16, 32]. 
}} 
\vspace{0.5cm}
\label{fig_pvmedium}
\end{figure}

\begin{figure}
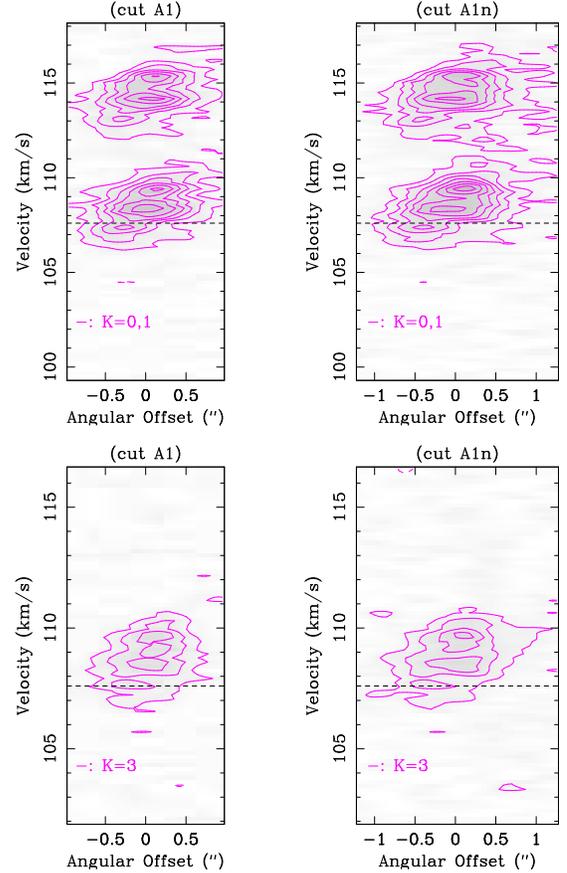


\hspace{0.25cm}
\begin{tabular}{ p{3.5cm} p{3.5cm}  }
\rotatebox{-90}{\includegraphics[scale=0.3]{pv_cuta1_k01.ps}} & \rotatebox{-90}{\includegraphics[scale=0.3]{pv_cuta1n_k01.ps}}  \\ 
\end{tabular}

\hspace{0.25cm}
\begin{tabular}{ p{3.5cm} p{3.5cm} }
\rotatebox{-90}{\includegraphics[scale=0.3]{pv_cuta1_k3.ps}} & \rotatebox{-90}{\includegraphics[scale=0.3]{pv_cuta1n_k3.ps}}  \\ 
\end{tabular}

\caption{\footnotesize{Position-Velocity (PV) diagram for the CH$_{3}$CN J=12-11, K=0,1,3 lines (magenta contour and grayscale) made in core A1. Dashed lines label the $v_{lsr}$=107.6 km\,s$^{-1}$. The PV cuts are drawn in Figure \ref{fig_moment2}, with diamonds indicating the end points of the cuts (i.e. positive angular offset). Magenta contours are 10.5 mJy\,beam$^{-1}$(3$\sigma$)$\times$[-1, 1, 2, 3, 4, 5, 6, 7, 8]. These PV diagrams of CH$_{3}$CN lines are generated from the Briggs robust=0 weighted image cubes, which has a synthesized beam of $\theta_{\mbox{\scriptsize{maj}}}$$\times$$\theta_{\mbox{\scriptsize{min}}}$=0$''$.73$\times$0$''$.51 (P.A.=77$^{\circ}$).
}} 
\vspace{0.5cm}
\label{fig_pvsmall}
\end{figure}

The comparison of the CH$_{3}$CN and the DCN PV diagrams show that the velocity offsets of core A1 and A2 relative to $v_{lsr}^{\mbox{\scriptsize{sys}}}$ are consistent with the global velocity gradient within the 0.3 pc radius in G33.92+0.11 A (Figure \ref{fig_pvmedium}).
These two massive cores are likely originated from the large scale accretion flow, and remain dynamically coupled to it.
In addition, on the $\sim$0.05 pc scale, we also observed the southeast-northwest velocity gradient within core A1.
From the velocity channel maps of the K=0 component, we can see the emission peak in core A1 moving from the southeast to the northwest, starting from the velocity channel of 106.18 km\,s$^{-1}$ to the velocity channel of 108.43 km\,s$^{-1}$ (Figure \ref{fig_channelch3cnk3}).
This motion is blended with a more redshifted component, which is moving from south to north.
The velocity channel map of the K=3 component trace better the south-north velocity gradient. 
We hypothesize that the south-north component is tracing the inner part of the mini-arm-A1n, although we cannot rule out the possibility that there is another redshifted core. 
The `cut A1' panels in Figure \ref{fig_pvsmall} show the PV diagrams of the K=0, 1, and 3 components of the CH$_{3}$CN J=12-11 line, which were generated from the PV cut closely following the southeast-northwest component in core A1 (see also Figure \ref{fig_moment2}, top right panel).
We also do not resolve the Keplerian velocity profile in core A1 from the CH$_{3}$CN PV diagrams (Figure \ref{fig_pvsmall}).
However, the highly accelerated gas in the center of core A1 may already be photoionized, and therefore cannot be detected in CH$_{3}$CN emission. 
The absence of high velocity CH$_{3}$CN emission may also because of chemical segregation (see also Jim{\'e}nez-Serra et al. 2012).
Away from the center of core A1, CH$_{3}$CN may not be excited or is chemically less enhanced due to the lower gas temperature.

We tentatively interpret the marginally resolved southeast-northwest velocity gradient by the rotational motion in A1.
If this is the case, the consistently detected southeast-northwest velocity gradients within core A1 and on the larger scale will be strongly indicative to a coherent gas accretion flow from large to small scale (more in Section \ref{sub:velocity}).
These hypotheses can be tested in future higher angular resolution observations.
We do not have enough sensitivity and angular resolution at the moment to resolve the velocity gradient within core A2 (and core A5).


\section{Discussion} \label{sec:discussion}
We derive gas mass distribution based on the dust continuum images, in Section \ref{sub:matter}.
In Section \ref{subsec:satellite}, we discuss the stability of the dense gas cores, which are likely the current or future sites of intermediate- or high-mass star-formation.
Our interpretation of the resolved velocity field is provided in Section \ref{sub:implication}, which is based on our knowledge of gas mass distribution.
We discuss the physical implication of our observational results in Section \ref{sub:implication}.

\subsection{Matter Distribution} \label{sub:matter}
The gas mass $M_{\mbox{\scriptsize{H$_{2}$}}}$ as well as column density in the extended structures can be approximated by the optically thin dust emission formula
\begin{equation}
M_{\mbox{\scriptsize{H$_{2}$}}} = \frac{2\lambda^{3}Ra\rho D^{2}}{3hcQ(\lambda)J(\lambda,T_{d})}S(\lambda),
\label{eq_mass}
\end{equation}
where $R$ is the gas-to-dust mass ratio, $a$ is the mean grain radius, $\rho$ is the mean grain density, $D$ is the distance to the target, $Q$($\lambda$)$\propto$$\lambda^{-\beta}$ is the grain emissivity, $T_{d}$ is the dust temperature, $S(\lambda)$ is the flux of the dust emission at the given wavelength, $J(\lambda, T_{d})=1/[\mbox{exp}(hc/\lambda k_{B}T_{d})-1]$ (Hildebrand 1983).
The $c$, $h$, and $k_{B}$ are the light speed, the Planck constant, and the Boltzmann constant, respectively. 
Following Lis at al. (1998), we adopt the standard values $R$=100, $a$=0.1 $\mu$m, $\rho$=3 g\,cm$^{-3}$, $Q$($\lambda$=350 $\mu$m)=1$\times$10$^{-4}$.
Based on the previous extensive NH$_{3}$ survey toward the high-mass star-forming regions (Lu et al. 2014), and our previous NH$_{3}$ rotational temperature measurements on this region (Liu et al. 2012b), we adopt a constant $T_{d}$=20 K on the large scale, and $T_{d}$$\sim$30 K in the inner 0.3 pc ($\sim$9$''$) radius of G33.92+0.11 A.

The gas mass of G33-north and G33-south in the polygon regions in Figure \ref{fig_continuum1} are 9700 $M_{\odot}$ and 5300 $M_{\odot}$, respectively.
The gas column density in these two regions are in the range of $N_{\mbox{\scriptsize{HI+H$_{2}$}}}$$\sim$2.5-7.1$\times$10$^{22}$ cm$^{-2}$, with a mean of $\sim$4.4$\times$10$^{22}$ cm$^{-2}$.
Given the assumed gas temperature and column density, and assuming that the width of the filaments are $\sim$1-2 pc, the corresponding thermal Jeans length and Jeans mass are 0.20--0.47 pc and 7.2-17 $M_{\odot}$, respectively; the local free-fall collapsing timescale $t_{ff}$ for the filaments is 0.42-1.0 Myrs.
With the resolution of SHARC2/CSO ($\sim$0.34 pc) and sensitivity (3$\sigma$ mass sensitivity $\sim$ 10.5 $M_{\odot}$, see Section \ref{subsec:cso}) we did not convincingly resolve the regularly spaced massive gas clumps in G33-north and G33-south (e.g. as seen in the more evolved OB cluster-forming region G10.6-0.4, see Liu et al. 2012a).  
Nevertheless, there are a few localized overintensities at Jeans mass scales. 
The gas structures western of G33.92+0.11 A are embedded within the localized gas clumps but are difficult to identify with the limited angular resolution of the SHARC2/CSO image.
The enclosed gas mass in the central $\sim$5 pc radius is $\sim$9.3$^{+2.7}_{-3.6}$$\times$10$^{4}$ $M_{\odot}$, where the error incorporates the uncertainties in dust temperature/opacity, as well as the foreground/background confusion.
The global free-fall collapsing timescale for the region inside a 5 pc radius is $\sim$1.1 Myrs.
We note that the detected gas mass within the central 5 pc radius in G33.92+0.11 is comparable to that in the Galactic mini-starburst region W49N (Galv\'{a}n-Madrid et al. 2013). 
Therefore, we think G33.92+0.11 has the potential of forming a very luminous OB cluster in the future. 
We also note that the central $\sim$0.3 pc radius massive clump G33.92+0.11 A is comparably massive ($\gtrsim$3500 $M_{\odot}$) as the $\sim$5 pc scale filaments G33-north/south, which indicates the high gas concentration in the cloud center.

Finally, we summarize this section with two resolved features in common, from large to small scales. 
First, we detect the brightest sources at the center of the large scale 350 $\mu$m map, and of the high angular resolution ALMA 1.3 mm image (Figures \ref{fig_continuum1}, \ref{fig_continuum2}). 
Within the 5 pc radius, the 350 $\mu$m image shows that $\sim$3.8\% of the gas mass is concentrated within 0.4\% of the projected area (i.e. G33.92+0.11 A). 
Within source A inside a 0.3 pc radius, ALMA resolved that $\sim$5.7\% of the gas mass is concentrated to 0.7\% of the projected area (i.e. core A1).
Such a (self-)similarity in fraction of gas mass concentrated in projected area may infer an anchored gravitational collapse from large to small scales. 
Although these massive gas structures (i.e. G33.92+0.11 A and core A1) do not dominate the gas mass in the referred areas, locally they can be viewed as the dominant compact gravitational source.
Both G33.92+0.11 A and core A1 have companions with comparably lower masses.
Second, we resolved the parsec scale molecular gas arms at where the $\sim$5 pc scale gas streams converge to the massive molecular clump.
We also resolved the molecular gas mini-arms, which connect the $<$0.1 pc dense molecular core with the parsec scale molecular arms.
The length of the observed molecular gas arms and mini-arms are $\sim$2--3 times the size of G33.92+0.11 A and core A1, respectively.

\subsection{Gravitational stability of the satellite cores}\label{subsec:satellite}
\begin{figure}
\hspace{-1.3cm}
\begin{tabular}{ p{8.5cm} }
\includegraphics[width=10cm]{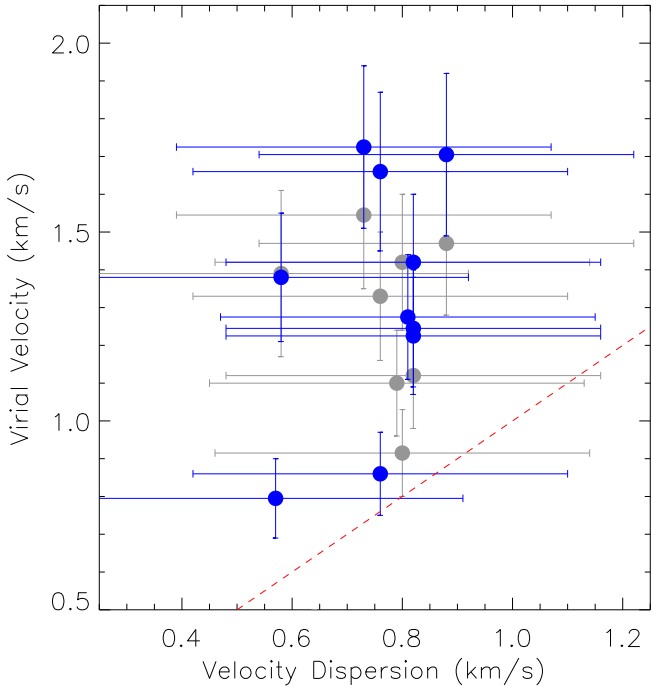} \\
\end{tabular}
\caption{\footnotesize{Estimated one-dimensional virial velocity dispersions ($v_{vir}$) and observed line-of-sight velocity dispersions ($\sigma_{v}$) of satellite cores A3, 4n, 4s, 5, 6, 9, 10, 11, 12 and 13 (Section \ref{subsec:satellite}). Blue symbols are the measurements based on the ALMA 12m+ACA Arrays 1.3 mm dust continuum image, and gray symbols are the measurements based on the ALMA 12m Array 1.3 mm dust continuum image. Red dashed line shows $v_{vir}$=$\sigma_{v}$. The horizontal error bars are $\pm$0.34 km\,s$^{-1}$ (i.e. half of the velocity channel spacing). The upper and lower bounds of the vertical error bars are derived by assuming the inverse-square radial density profile and the uniform density distribution, respectively. We note that core 12 is seriously confused with the negative sidelobes in the 12m Array alone image, and therefore its flux cannot be measured. 
}} 
\vspace{0.3cm}
\label{fig_virial}
\end{figure}

To examine whether the observed satellite cores in molecular arms can collapse to form high-mass stars, we derive one-dimensional virial velocity dispersions 
\begin{equation}
v_{vir} = \sqrt{\frac{\alpha MG}{5R_{eff}}},
\label{eq_virial}
\end{equation}
for cores A3, A4n, A4s, A5--6, and A9--13 based on the formulation in Williams et al. (1994), and compare with the line-of-sight velocity dispersions ($\sigma_{v}$) traced by DCN 3-2. 
In Equation \ref{eq_virial}, the parameter $\alpha$ is the geometric factor which equals to unity for a uniform density profile and $5/3$ for an inverse square profile, $M$ is the gas mass, $G$ is the gravitational constant, and $R_{eff}$ is the effective radius of the dense clump.
We suggest that the DCN 3-2 line may be a particularly good dense gas tracer for this purpose, since DCN can survive 10$^{4-5}$ years after the onset of (proto)stellar feedback (Rodgers \& Millar 1996).
To measure core masses and sizes, we fit 2-dimensional Gaussians to the 1.3 mm dust continuum images (Figure \ref{fig_continuum2}) using the CASA task {\tt imfit}, and then estimate gas mass based on Equation 1. 
We assume the dust temperature $T_{d}$$\sim$30 K, and adopt $\beta$=1. 
To take into account of the effects of ambient gas contamination and the missing flux, we perform Gaussian fittings to both the ALMA 12m+ACA Array 1.3 mm dust continuum image, and the dust continuum image generated from the 12m Array data alone.
For each of the cores, we extract the DCN line spectrum in the area defined by the 2-dimensional Gaussian fitting, and then perform 1-dimensional Gaussian fitting to the extracted spectrum to measure $\sigma_{v}$.
The results are presented in Figure \ref{fig_virial}.

We found that the derived $v_{vir}$ is in general larger than $\sigma_{v}$.
This may imply that (1) the core masses are significantly overestimated by a factor of $\sim$2.5 or more, or (2) the gas motions do not provide sufficient support against the local gravitational collapse, although we cannot rule out the case (3): all cores are rotationally flattened and the dominant gas motions are perpendicular to the line-of-sight.
Case (1) may be true if large dust grains are present in the satellite cores, such that locally the value of $\beta$ is close to 0.
Another option is that actual gas-to-dust mass ratio is a few times lower than the assumed value (100). 
Our data do not allow discriminating these cases.
The values of $\beta$ in satellite cores can be measured by performing dust continuum observations at other wavelengths (e.g. 0.85mm or 3 mm).
Case (3) needs to be tested by future higher angular resolution observations. 
We tentatively favor case (2), or a combination of cases (2) and (3) for some cores, since there were localized high velocity CO 2-1 outflow features detected in the previous observations (Liu et al. 2012b).
In any case, the evidence makes unlikely that the supersonic gas motions in the satellite cores are much faster than the actual $v_{vir}$, such that these cores are not stable.

\subsection{Velocity Fields} \label{sub:velocity}
The observed gas motions in G33.92+0.11 A, in particular, in arms-c1, c2, c4 and c5, may be explained by rotational and free-fall motions, accelerated by the enclosed matter distribution.
For example, we consider that the enclosed mass $M(r)$ within the radius $r$ can be approximated by $M(r)=M_{0}\cdot r^{n}$.
Surrounding the $\sim$0.3 pc scale flattened rotating massive molecular clump, the value of $n$ may be in between 1--3 (c.f. the solution for a collapsing slowly rotating isothermal cloud given by Terebey et al. 1984). 
The gas velocity can be estimated by $v_{\mbox{\scriptsize{free-fall}}}$$\sim$$\sqrt{2GM_{0}}$$\cdot$$r^{(n-1)/2}$.
Based on the observed increasing velocity offset at larger radii, the value of $n$ may be approximately in between 2--3, which is expected in the region where the large scale gas streams are converging to the flattened rotating structure. 
The PV `cut c4a' does not show a clear velocity gradient in angular offsets of -3$''$--0$''$.5 (Figure \ref{fig_pvlarge}), which is likely because the orientation of this PV cut is perpendicular to this velocity gradient of the rotating clump (c.f. Figure \ref{fig_moment1} and \ref{fig_moment2}). 
At larger angular offset, the `cut c4a' detects the exterior redshifted infall (more below).
We note that this analysis is preliminary because we do not know exactly how the inclination of the individual molecular arms vary with position.
Defining the matter distribution as $M(r)$ on the $<$0.3 pc scale, and deprojecting the observed line-of-sight velocity, are not possible because of this reason. 
Due to the complex morphology, the matter distribution may also be poorly approximated by $M(r)$. 
The center of mass of G33.92+0.11 A is likely located in between cores A1 and A2, but cannot be precisely determined at this moment.
In addition, our observations do not allow us to separate the rotational motion from infall. 
Nevertheless, qualitatively we think our interpretation is physically plausible and has to be considered in other case studies.

On smaller scales, the observed projected velocity surrounding core A1 is $\sim\pm$1.00 km\,s$^{-1}$ relative to its systemic velocity $v_{lsr}$=108.47 km\,s$^{-1}$, at the 6.8$\pm$0.8$\times$10$^{-3}$ pc projected radius.
Assuming that the embedded 100-300 $M_{\odot}$ gas mass (Liu et al. 2012b) in the core A1 is dominant, and assuming that the surrounding gas in the $<$0.01 pc radius follows the nearly circular orbits, the rotating speed $v_{R}(\mbox{0.0068 pc})$ can be estimated by $v_{R} = [(G\cdot\mbox{(100-300 $M_{\odot}$}))/(\mbox{0.0068 pc})]^{1/2}=$7.7-13 km\,s$^{-1}$.
We therefore can interpret the southeast-northwest velocity gradient around A1 by the rotation inclined by 4$^{\circ}$-7$^{\circ}$ along the position angle of $\sim$45$^{\circ}$.
The CH$_{3}$CN velocity field surrounding the core A2 is more complicated, in particular, in the mini-arm-Aw.
Nevertheless, we found that at the 0.035-0.05 pc position offset from core A2, the observed $\sim$0.4 km\,s$^{-1}$ velocity offset of the mini-arm-A2w from the centroid velocity of core A2, can be consistently interpreted by the gravitationally accelerated motion by core A2 (and A1), projected to a similar inclination angle to that of core A1.

If the velocity gradient on the $>$0.3 pc scale is consistent with that on the $<$0.3 pc as well as the southeast-northwest velocity gradient within core A1, we could expect the entire molecular arm-S1 and S2 (i.e. southern than decl. (J2000)$\sim$00$^{\circ}$55$'$23$''$) to be blue shifted.
We present the PV `cut c4b', `cut c4c', `cut c4d' in Figure \ref{fig_pvlarge}, which trace the gas motions in the molecular arm-S1 and S2.
We instead found that the line-of-sight motion of these two arms converge smoothly from the redshifted velocity in the north to the blueshifted velocity in the south and southeast.
These results are consistent with the previous 0.6 km\,s$^{-1}$ velocity resolution observations of the NH$_{3}$ (1,1) hyperfine inversion lines (Liu et al. 2012b).   
In the marginally spatially resolved system, this approximately linear velocity gradient is most commonly interpreted by the rotational motion of a toroid. 
In the particular case of G33.92+0.11, we are motivated by the resolved matter distribution, and the morphology of the arms, to alternatively consider that the parsec scale velocity gradient in the north-south direction can be the free-fall dominant motion following the eccentric orbits.
The resolved velocity gradient from the molecular arm-c4 to arms-S1, S2, is qualitatively similar to the numerical simulation for the infalling filament connected with the low-mass cluster-forming region in Bonnell et al.(2008).
The southern molecular arm may be gravitationally accelerated by the massive core A1, and therefore turns into the redshifted velocity in regions close to A1.
From both the average velocity map and the PV diagrams of $^{13}$CS 5-4 (Figure \ref{fig_moment1}, \ref{fig_pvlarge}), we observe the smaller velocity scale outside the 0.3 pc radius of G33.92+0.11 A, than the velocity of the gas arms interior to it. 
This can be understood if the observed gas structures within the 0.3 pc radius of G33.92+0.11 A dominate the enclosed mass within the $\sim$1 pc radius.

\begin{figure}
\hspace{-0.2cm}
\begin{tabular}{ p{8.5cm} }
\includegraphics[width=8.5cm]{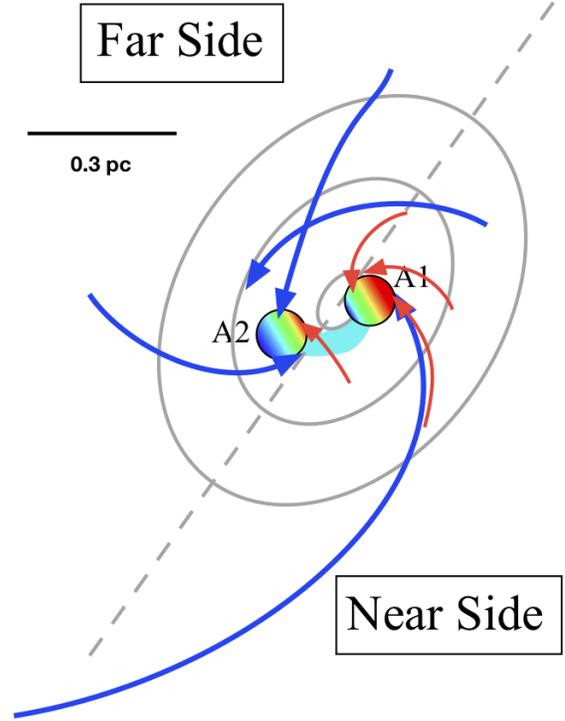} \\
\end{tabular}
\caption{\footnotesize{A schematic model of the resolved region. Gray ellipses are the projected iso-radius contours of an inclined circular rotating plane. The near and far sides of this inclined system are separated by the dashed line. The blue and red arrows show the spiraling $\sim$0.5 pc scale molecular arms, and the mini-arms connected with cores A1 and A2, respectively. The filled circles are the cores A1 and A2, with color scales indicating the line-of-sight velocity gradients caused by their spinning motion. These two massive cores are rotating about each other, such that the core A1 is redshifted with respect to core A2. Observationally, the velocity distribution in A2 is confused with the blueshifted tidally interacting feature connecting from A2 to A1, which is indicated as the filled cyan area. We note that some of the arms can be off plane.
}} 
\vspace{0.3cm}
\label{fig_model}
\end{figure}

We summarize our working hypothesis to interpret the resolved structures and kinematics in the schematics model in Figure \ref{fig_model}.
We think the molecular gas streams are converging from the larger than 1 pc radius towards the center of G33.92+0.11 A, due to gravitational infall acceleration.
The gas motions become dominated by the rotational motion on the $\sim$0.3 pc scale.
Within the 0.3 pc scale radius, we detect the gas mini-arms connecting to the central core A1 and A2, which are rotating about each other. 
These gas mini-arms show $\lesssim$1 km\,s$^{-1}$ velocity offsets from the ambient dense gas, which can be explained by the gravitational acceleration and the tidal interaction of the dense core.
However, we cannot yet rule out that the motion of the gas mini-arms is powered by the expansional motion of the ionized gas, or the molecular wind/jet, which needs to be examined in higher resolution observations.

\subsection{Physical Implications} \label{sub:implication}
The magnetic field in rapidly collapsing molecular clumps is presumably weak. 
The initial turbulence is rapidly dissipated in such dense regions, although can be replenished by the protostellar turbulence in a later evolutionary stage (Wang et al. 2010).
The rotational motion can play an important role in supporting collapsing molecular clumps, which permits localized fragmentation and star-formation in longer timescales. 
It is not yet clear how to transport angular momentum outward efficiently on spatial scale of a fraction of a parsec.
The excess of angular momentum may leave its footprint on the morphology of the molecular clumps, and may therefore bias the upper end of the core/stellar mass function.

Observationally, we have found clumpy rotating toroids and spiral-like interacting gas features toward the low-mass cluster-forming region L1287 (Juar\'{e}z et al. in prep.), the OB cluster-forming molecular clumps (e.g. G10.6-0.4: Liu et al. 2010; G35: S{\'a}nchez-Monge et al. 2013, Qiu et al. 2013; NGC6334\,V: Juar\'{e}z et al. in prep.), and the Galactic mini-starburst region W49N (Welch et al.1987, Galv\'{a}n-Madrid et al. 2013, etc).
These results imply that this physical problem is spatially scalable, and the previously spatially resolved systems are the large-scale examples seen in a closer to face-on projection.
However, toroids may also be the projected central OB cluster-forming cores surrounded by satellite cores. 
The critical spatial scales to look at, are the centrifugal radius (or radii), where the centrifugal force marginally balances the gravitational force from the enclosed molecular and stellar mass. 
We expect more of these systems to be discovered or be spatially resolved in high angular resolution ALMA observations.
We hypothesize that the inward migration of satellite cores can explain the formation of the extremely concentrated dense condensations, similar to what was resolved in NH$_{3}$ (3,3) satellite hyperfine line absorption against the UC H\textsc{ii} region G10.6-0.4 (Sollins \& Ho 2005; Liu et al. 2010).

Finally, we hypothesize that the spiraling arm-like asymmetry is essential in the gravitationally dominated accretion flow, and needs to be considered in modeling frameworks. 
These structures may either be induced at regions where the eccentric accretion streams collide with each other, or be created by tidal interactions. 
We note that tidal accretion of eccentric accretion streams, or tidally interacting hot cores, may deposit the hot molecular gas tracers from the hot cores to the entire parsec scale massive molecular clump. 

We would like to emphasize that the location of the objects in the region discussed favor the idea of the convergence of parsec scale gas accretion flows onto a central clump.
Our high spectral resolution DCN 3-2 and $^{13}$CS 5-4 line observations suggest that this massive molecular clump is fed by free-falling exterior gas streams (or arms).
Based on the resolved images, we found that accretion flows in this region are highly dynamical and spatially non-uniform.
Individual accretion gas streams likely carry different amount of specific angular momentum. 
Even in the case that the averaged specific angular momentum on the large scale is negligibly small, it remains uncertain how the specific angular momentum carried by individual gas streams will alter the gas accretion and fragmentation. 
High angular resolution molecular line observations to study a large ensemble may be required to elucidate this issue.

\section{Summary} \label{sec:summary}
We observed the approximately face-on young OB cluster-forming region G33.92+0.11, using SHARC2/CSO, {\it Herschel}, and ALMA.
On the $>$1 pc scale, the SHARC2 0.35 mm dust continuum image trace the filamentary gas streams converging to the central $\sim$10$^{4}$ $M_{\odot}$ massive molecular gas clump G33.92+0.11 A.
The high resolution ALMA 1.3 mm dust continuum images reveal abundant dense satellite cores, which are closely associated with the $\sim$0.5 pc scale molecular gas arms orbiting the two central highest mass cores.
We found that the DCN 3-2 line emission correlates well with the 1.3 mm dust continuum emission, while the $^{13}$CS 5-4 line shows further enhanced emission towards the gas mini-arms connecting with the central massive cores and the potentially shock heated gas.
These two lines consistently trace a southeast-northwest global velocity gradient within the 0.3 pc radius in G33.92+0.11 A, which may be interpreted by the rotational and infall motion.
The small line width and velocity differences between dense features strongly supports the interpretation of a flattened object viewed nearly face-on. 
The spiral arms would indicate that the flattening is mostly likely due to rotation, although the full rotation speed is hard to resolve because of the face-on geometry.
The CH$_{3}$CN lines appear to trace the same sense of rotation to the $<$0.05 pc scale hot core regions, which indicated that there may be a coherent accretion flow continuing from large to small scale.
The parsec scale molecular arm-S1, S2 traced by the 1.3 mm dust continuum emission, and the molecular line emission, may be the infall gas streams feeding massive molecular clump G33.92+0.11 A.
We highlight the resolved arm-like features connecting with the massive localized structures, and hypothesize  that the arm-like morphology is essential in star and cluster forming regions of all masses. 
We derived higher virial velocity dispersions than observed velocity dispersions for the satellite molecular cores embedded in molecular arms, which suggest that these cores are gravitationally unstable, and may collapse to form high-mass stars.

\vspace{-0.3cm}
\acknowledgements
This paper makes use of the following ALMA data: ADS/JAO.ALMA 2012.1.00387.S. 
ALMA is a partnership of ESO (representing its member states), NSF (USA) and NINS (Japan), together with NRC (Canada) and NSC and ASIAA (Taiwan), in cooperation with the Republic of Chile. 
The Joint ALMA Observatory is operated by ESO, AUI/NRAO and NAOJ.
This material is based upon work at the Caltech Submillimeter Observatory, which is operated by the California Institute of Technology.
In the present paper we discussed observations performed with the ESA Herschel Space Observatory (Pilbratt et al. 2010), in particular employing Herschel's large telescope and powerful science payload to do photometry using the PACS (Poglitsch et al. 2010) and SPIRE (Griffin et al. 2010) instruments. 
Herschel is an ESA space observatory with science instruments provided by European-led Principal Investigator consortia and with important participation from NASA.
This research made use of Montage, funded by the National Aeronautics and Space Administration's Earth Science Technology Office, Computation Technologies Project, under Cooperative Agreement Number NCC5-626 between NASA and the California Institute of Technology. 
Montage is maintained by the NASA/IPAC Infrared Science Archive.
HBL thanks Simon Radford and other CSO staffs for assisting the operation of SHARC2.
HBL thanks ASIAA for supporting series of related researches.
HBL acknowledge funding under MOST grant number 103-2119-M-001-010-MY2.
I.J.-S. acknowledges funding from the People Programme (Marie Curie Actions) of the European Union's Seventh Framework Programme (FP7/2007-2013) under REA grant agreement number PIIF-GA-2011-301538.
CRZ acknowledges support from program CONACYT CB2010-152160, Mexico. 

{\it Facilities:} \facility{ALMA, SHARC2/CSO, Herschel}



\appendix 

\section{A. Identified Lines from the ALMA 12m + ACA Observations}
We present the observed molecular lines in the central 10$''$ region in G33.92+0.11 A, in Table \ref{tab-lines}.

\begin{table*}[h]

\caption{Molecular Lines observed within our Cycle-1 ALMA data}
\label{tab-lines}
\hspace{3.5cm}
\vspace{-3cm}
\begin{tabular}{cccccccc}\hline\hline
Species & Transition & Frequency & Eu \\
		   &                & (MHz)       & (K) \\\hline
		   
c-C$_3$H$_2$ & 3$_{3,0}$$\rightarrow$2$_{2,1}$ & 216278.76 & 19.5 \\ 

CCD & N=3-2, J=7/2-5/2, F=9/2-7/2 & 216372.83 & 20.8 \\
    & N=3-2, J=7/2-5/2, F=7/2-5/2\tablenotemark{a} & 216373.32 & 20.8 \\
    & N=3-2, J=5/2-3/2, F=7/2-5/2\tablenotemark{b} & 216428.32 & 20.8 \\
    & N=3-2, J=5/2-3/2, F=3/2-1/2 & 216428.76 & 20.8 \\

CH$_3$CHO & 11$_{1,10}\rightarrow$10$_{1,9}$ E & 216581.94 & 64.9 \\ 
          & 11$_{1,10}\rightarrow$10$_{1,9}$ A & 216630.22 & 64.8 \\

HDCS & 	7$_{0,7}\rightarrow$6$_{0,6}$ &	216662.43 & 41.6 \\
     &  7$_{2,6}\rightarrow$6$_{2,5}$ & 216931.37 & 77.6 \\
     &  7$_{2,5}\rightarrow$6$_{2,4}$ & 217263.69 & 77.6 \\

H$_2$S &  2$_{2,0}\rightarrow$2$_{1,1}$ & 216710.44 & 84.0 \\

CH$_3$OH & 5$_{1,4}\rightarrow$4$_{2,2}$ & 216945.60 & 55.9 \\

SiO  & 5$\rightarrow$4 & 217104.98 & 31.3 \\

DCN & 3$\rightarrow$2 & 217238.63 & 20.9 \\

$^{13}$CN &  N=2-1, J=3/2-1/2, F1=1-1, F=2-1 & 217072.80 & 15.7 \\
          &  N=2-1, J=3/2-1/2, F1=1-1, F=2-2 & 217074.24 & 15.7 \\
          &  N=2-1, J=3/2-1/2, F1=1-0, F=0-1 & 217264.64 & 15.7 \\
          &  N=2-1, J=3/2-1/2, F1=1-0, F=1-1 & 217277.68 & 15.7 \\
          &  N=2-1, J=5/2-3/2, F1=2-1, F=2-2 & 217286.80 & 15.7 \\
          &  N=2-1, J=5/2-3/2, F1=2-1, F=1-1 & 217290.82 & 15.7 \\
          &  N=2-1, J=5/2-3/2, F1=2-1, F=1-0 & 217296.60 & 15.7 \\
          &  N=2-1, J=3/2-3/2, F1=2-2, F=2-2 & 217298.94 & 15.7 \\
          &  N=2-1, J=5/2-3/2, F1=2-1, F=2-1 & 217301.18 & 15.7 \\
          &  N=2-1, J=5/2-3/2, F1=2-1, F=3-2 & 217303.19 & 15.7 \\
          &  N=2-1, J=3/2-1/2, F1=1-0, F=2-1 & 217304.93 & 15.7 \\
          &  N=2-1, J=3/2-3/2, F1=2-2, F=3-3 & 217306.12 & 15.7 \\
          &  N=2-1, J=5/2-3/2, F1=2-1, F=3-2 & 217428.56 & 15.7 \\
          &  N=2-1, J=5/2-3/2, F1=2-1, F=2-1 & 217436.35 & 15.7 \\
          &  N=2-1, J=5/2-3/2, F1=2-1, F=2-2 & 217437.70 & 15.7 \\
          &  N=2-1, J=3/2-1/2, F1=2-1, F=2-2 & 217437.82 & 15.7 \\
          &  N=2-1, J=3/2-1/2, F1=2-1, F=1-1 & 217443.72 & 15.7 \\
          &  N=2-1, J=5/2-3/2, F1=3-2, F=4-3 & 217467.15 & 15.7 \\
          &  N=2-1, J=5/2-3/2, F1=3-2, F=2-1 & 217469.15 & 15.7 \\
          &  N=2-1, J=5/2-3/2, F1=3-2, F=2-2 & 217480.56 & 15.7 \\
          &  N=2-1, J=5/2-3/2, F1=3-2, F=3-3 & 217483.61 & 15.7 \\

t-C$_2$H$_5$OH & 5$_{3,3}\rightarrow$4$_{2,2}$ & 217803.69 & 23.9 \\

c-HCCCH & 6$_{1,6}\rightarrow$5$_{0,5}$\tablenotemark{c} & 217822.15 & 38.6 \\ 
        & 5$_{1,4}\rightarrow$4$_{2,3}$ & 217940.05 & 35.4 \\

CH$_3$CN & 12$\rightarrow$11 K=4 & 220679.29 & 183.1 \\
         & 12$\rightarrow$11 K=3 & 220709.02 & 133.2 \\
         & 12$\rightarrow$11 K=2 & 220730.26 & 97.4 \\
         & 12$\rightarrow$11 K=1 & 220743.01 & 76.0 \\
         & 12$\rightarrow$11 K=0 & 220747.26 & 68.9 \\

OCS & 19$\rightarrow$18 & 231060.99 & 110.9 \\

$^{13}$CS & 5$\rightarrow$4 & 231220.69 & 33.3 \\

H$\alpha$ &  $n$=30 & 231900.93 & 0.0 \\

He$\alpha$ & $n$=30 & 231995.43 & 0.0 \\

CH$_3$OCH$_3$ & 13$_{0,13}\rightarrow$12$_{1,12}$ EE\tablenotemark{d} & 231987.82 & 80.9 \\

H$_2$C$^{34}$S & 7$_{1,7}\rightarrow$6$_{1,6}$ & 232754.71 & 57.9 \\\hline

\end{tabular}
\vspace{3cm}
\caption*{\footnotesize{
\hspace{1.4cm}
\begin{tabular}{ p{12cm} }
$^{a}$\,Blended with transition CCD N=3-2, J=7/2-5/2, F=5/2-3/2. \\
$^{b}$\,Blended with transition CCD N=3-2, J=5/2-3/2, F=5/2-3/2. \\
$^{c}$\,Blended with transition c-HCCCH 6$_{0,6}\rightarrow$5$_{1,5}$. \\
$^{d}$\,The CH$_3$OCH$_3$ 13$_{0,13}\rightarrow$12$_{1,12}$ transitions from the AA, AE and EA species are blended with this line (frequencies are within $\pm$0.2 MHz).\\ 
\end{tabular}
}}
\end{table*}


\begin{thebibliography}{}

\bibitem[Araya et al.(2005)]{2005ApJS..157..279A} Araya, E., Hofner, P., Kurtz, S., Bronfman, L., \& DeDeo, S.\ 2005, \apjs, 157, 279

\bibitem[Bonnell et al.(2008)]{2008MNRAS.389.1556B} Bonnell, I.~A., Clark, 
P., \& Bate, M.~R.\ 2008, \mnras, 389, 1556

\bibitem[Burkert \& Hartmann(2004)]{2004ApJ...616..288B} Burkert, A., \& Hartmann, L.\ 2004, \apj, 616, 288

\bibitem[Busquet et al.(2013)]{2013ApJ...764L..26B} Busquet, G., Zhang, Q., Palau, A., et al.\ 2013, \apjl, 764, L26

\bibitem[Cesaroni et al.(2011)]{2011A&A...533A..73C} Cesaroni, R., Beltr{\'a}n, M.~T., Zhang, Q., Beuther, H., \& Fallscheer, C.\ 2011, \aap, 533, A73

\bibitem[Chen \& Ostriker(2014)]{2014ApJ...785...69C} Chen, C.-Y., \& Ostriker, E.~C.\ 2014, \apj, 785, 69 

\bibitem[Churchwell(2002)]{2002ARA&A..40...27C} Churchwell, E.\ 2002, \araa, 40, 27

\bibitem[De Pree et al.(1997)]{1997ApJ...482..307D} De Pree, C.~G., Mehringer, D.~M., \& Goss, W.~M.\ 1997, \apj, 482, 307

\bibitem[Dobbs et al.(2013)]{2013arXiv1312.3223D} Dobbs, C.~L., Krumholz, M.~R., Ballesteros-Paredes, J., et al.\ 2013, arXiv:1312.3223

\bibitem[Dowell et al.(2003)]{2003SPIE.4855...73D} Dowell, C.~D., Allen, C.~A., Babu, R.~S., et al.\ 2003, \procspie, 4855, 73

\bibitem[Fish et al.(2003)]{2003ApJ...587..701F} Fish, V.~L., Reid, M.~J., Wilner, D.~J., \& Churchwell, E.\ 2003, \apj, 587, 701 

\bibitem[Galv{\'a}n-Madrid et al.(2009)]{2009ApJ...706.1036G} Galv{\'a}n-Madrid, R., Keto, E., Zhang, Q., Kurtz, S., Rodr{\'{\i}}guez, L.~F., \& Ho, P.~T.~P.\ 2009, \apj, 706, 1036


\bibitem[Galv{\'a}n-Madrid et al.(2013)]{2013ApJ...779..121G} Galv{\'a}n-Madrid, R., Liu, H.~B., Zhang, Z.-Y., et al.\ 2013, \apj, 779, 121 

\bibitem[{Griffin {et~al.}(2010)Griffin, et al.}]{Griffin:2010hz} Griffin, M.~J., Abergel, A., Abreu, et al.\ 2010, A\&A, 518, L3

\bibitem[Hartmann et al.(2012)]{2012MNRAS.420.1457H} Hartmann, L., Ballesteros-Paredes, J., \& Heitsch, F.\ 2012, \mnras, 420, 1457 

\bibitem[Hatchell et 
al.(1998)]{1998A&A...332..695H} Hatchell, J., Millar, T.~J., \& Rodgers, S.~D.\ 1998, \aap, 332, 695

\bibitem[Hildebrand(1983)]{1983QJRAS..24..267H} Hildebrand, R.~H.\ 1983, \qjras, 24, 267

\bibitem[Howard et al.(2014)]{2014MNRAS.438.1305H} Howard, C.~S., Pudritz, R.~E., \& Harris, W.~E.\ 2014, \mnras, 438, 1305 


\bibitem[Jim{\'e}nez-Serra et al.(2012)]{2012ApJ...753...34J} 
Jim{\'e}nez-Serra, I., Zhang, Q., Viti, S., Mart{\'{\i}}n-Pintado, J., 
\& de Wit, W.-J.\ 2012, \apj, 753, 34 



 

\bibitem[Keto et al.(1991)]{1991ApJ...383..639K} Keto, E.~R., Lattanzio, 
J.~C., \& Monaghan, J.~J.\ 1991, \apj, 383, 639

\bibitem[Kov{\'a}cs(2008)]{2008SPIE.7020E..45K} Kov{\'a}cs, A.\ 2008, 
\procspie, 7020, 45


\bibitem[Lis et al.(1998)]{1998ApJ...509..299L} Lis, D.~C., Serabyn, E., Keene, J., Dowell, C.~D., Benford, D.~J., Phillips, T.~G., Hunter, T.~R., \& Wang, N.\ 1998, \apj, 509, 299 

\bibitem[Liu et al.(2010)]{2010ApJ...722..262L} Liu, H. B., Ho,  P.~T.~P., Zhang, Q., Keto, E., Wu, J., \& Li, H.\ 2010, \apj, 722, 262



\bibitem[Liu et al.(2012)]{2012ApJ...745..61L} Liu, H.~B., Quintana-Lacaci, G., Wang, K., et al.\ 2012a, \apj, 745, 61 

\bibitem[Liu et al.(2012)]{2012ApJ...756...10L} Liu, H.~B., Jim{\'e}nez-Serra, I., Ho, P.~T.~P., et al.\ 2012b, \apj, 756, 10 

\bibitem[Liu et al.(2012)]{2012thesis} Liu, H.~B. 2012c, PhD thesis, National Taiwan University

\bibitem[Liu et al.(2012)]{2012ApJ...756..195L} Liu, H.~B., Hsieh, P.-Y., 
Ho, P.~T.~P., et al.\ 2012d, \apj, 756, 195

\bibitem[Liu et al.(2013)]{2013ApJ...771...71L} Liu, H.~B., Qiu, K., Zhang, Q., Girart, J.~M., \& Ho, P.~T.~P.\ 2013, \apj, 771, 71 

\bibitem[Lu et al.(2014)]{2014arXiv1405.7933L} Lu, X., Zhang, Q., Liu, 
H.~B., Wang, J., \& Gu, Q.\ 2014, arXiv:1405.7933



\bibitem[Mackay(1999)]{1999MNRAS.304...61M} Mackay, D.~D.~S.\ 1999, \mnras, 304, 61


\bibitem[McMullin et al.(2007)]{2007ASPC..376..127M} McMullin, J.~P., Waters, B., Schiebel, D., Young, W., \& Golap, K.\ 2007, Astronomical Data Analysis Software and Systems XVI, 376, 127


\bibitem[Myers(2009)]{2009ApJ...700.1609M} Myers, P.~C.\ 2009, \apj, 700, 1609 

\bibitem[Myers(2011)]{2011ApJ...735...82M} Myers, P.~C.\ 2011, \apj, 735, 82



\bibitem[{Pilbratt {et~al.}(2010)Pilbratt, Riedinger, Passvogel, Crone, Doyle, Gageur, Heras, Jewell, Metcalfe, Ott, \& Schmidt}]{Pilbratt:2010en} Pilbratt, G.~L., Riedinger, J.~R., Passvogel, T., et al.\ 2010, A\&A, 518, L1

\bibitem[{Poglitsch {et~al.}(2010)Poglitsch et al}]{Poglitsch:2010bm}Poglitsch, A., Waelkens, C., Geis, N. et al.\ 2010,\aa, 518, L2

\bibitem[Qiu et al.(2013)]{2013ApJ...779..182Q} Qiu, K., Zhang, Q., Menten, K.~M., Liu, H.~B., \& Tang, Y.-W.\ 2013, \apj, 779, 182


\bibitem[Rodgers \& Millar(1996)]{1996MNRAS.280.1046R} Rodgers, S.~D., \& Millar, T.~J.\ 1996, \mnras, 280, 1046

\bibitem[S{\'a}nchez-Monge et al.(2013)]{2013A&A...552L..10S} S{\'a}nchez-Monge, {\'A}., Cesaroni, R., Beltr{\'a}n, M.~T., et al.\ 2013, \aap, 552, L10

\bibitem[Sollins \& Ho(2005)]{2005ApJ...630..987S} Sollins, P.~K., \& Ho, P.~T.~P.\ 2005, \apj, 630, 987


\bibitem[Sault et al.(1995)]{1995ASPC...77..433S} Sault, R.~J., Teuben, P.~J., \& Wright, M.~C.~H.\ 1995, Astronomical Data Analysis Software and Systems IV, 77, 433


\bibitem[Schneider et al.(2010)]{2010A&A...520A..49S} Schneider, N., Csengeri, T., Bontemps, S., et al.\ 2010, \aap, 520, A49


\bibitem[Sutton et al.(1986)]{1986ApJ...311..921S} Sutton, E.~C., Blake, 
G.~A., Genzel, R., Masson, C.~R., \& Phillips, T.~G.\ 1986, \apj, 311, 921

\bibitem[Tan et al.(2014)]{2014arXiv1402.0919T} Tan, J.~C., Beltran, M.~T., 
Caselli, P., et al.\ 2014, arXiv:1402.0919

\bibitem[Terebey et al.(1984)]{1984ApJ...286..529T} Terebey, S., Shu, F.~H., \& Cassen, P.\ 1984, \apj, 286, 529

\bibitem[Toal{\'a} et al.(2012)]{2012ApJ...744..190T} Toal{\'a}, J.~A., V{\'a}zquez-Semadeni, E., \& G{\'o}mez, G.~C.\ 2012, \apj, 744, 190


\bibitem[Van Loo et al.(2014)]{2014ApJ...789...37V} Van Loo, S., Keto, E., \& Zhang, Q.\ 2014, \apj, 789, 37

\bibitem[V{\'a}zquez-Semadeni et al.(2007)]{2007ApJ...657..870V} V{\'a}zquez-Semadeni, E., G{\'o}mez, G.~C., Jappsen, A.~K., et al.\ 2007, \apj, 657, 870


\bibitem[Wang et al.(2010)]{2010ApJ...709...27W} Wang, P., Li, Z.-Y., Abel, T., \& Nakamura, F.\ 2010, \apj, 709, 27


\bibitem[Watt \& Mundy(1999)]{1999ApJS..125..143W} Watt, S., \& Mundy, L.~G.\ 1999, \apjs, 125, 143

\bibitem[Welch et al.(1987)]{1987Sci...238.1550W} Welch, W.~J., Dreher, J.~W., Jackson, J.~M., Terebey, S., \& Vogel, S.~N.\ 1987, Science, 238, 1550 

\bibitem[Williams et al.(1994)]{1994ApJ...428..693W} Williams, J.~P., de 
Geus, E.~J., \& Blitz, L.\ 1994, \apj, 428, 693 




\bibitem[Zhang et al.(2009)]{2009ApJ...696..268Z} Zhang, Q., Wang, Y.,  Pillai, T., \& Rathborne, J.\ 2009, \apj, 696, 268


\end{thebibliography}
\end{document}